\documentclass[twocolumn,showpacs,showkeys,preprintnumbers]{revtex4}
\usepackage{graphicx}
\usepackage{subfigure}
\usepackage{lscape}
\usepackage{dcolumn}
\usepackage{bm}
\usepackage{color}

\usepackage{amsmath} 
\usepackage{amssymb} 
\usepackage{multirow}
\usepackage{epstopdf}

\begin{document}
\title{A new basic air shower observable sensitive to the cosmic-ray elemental mass}
\author{A. Basak}
\email{ab.astrophysics@rediffmail.com}
\author{R. K. Dey}\thanks{Corresponding author}
\email{rkdey2007phy@rediffmail.com}
\affiliation{Department of Physics, University of North Bengal, Siliguri, WB 734 013 India}

\begin{abstract}
Based on a Monte Carlo simulation study of vertical extensive air showers (EAS) at the KASCADE location we introduce a new simple observable $\eta_{\rho{(45;310)}}$ (in short $\eta_{\rho}$) - the ratio between two lateral electron densities of an EAS measured at two well-defined radial distances indicated by the characteristic radial feature of the local age parameter (LAP). Our analyses of simulated data generated by the Group I nuclei with cosmic-ray elemental masses, $A_{\text{actual}}=1,4,12,16,24,32,40,56$, observed a double correlation of $\eta_{\rho}$ with, (i) the lateral shower age $s_{\text{av}}$, and (ii) $A_{\text{actual}}$ of the EAS initiating primary particle. Applying the first correlation to a new set of simulated showers initiated by the Group II nuclei with $A_{\text{actual}}=7,14,20,28,40,52$, the average difference between the lateral shower age $s_{\text{av}}$ obtained directly from the conventional LAP approach and the estimated one $s_{\text{est}}$ is found to be close to $0$. The second correlation involving $\eta_{\rho}$ with $A_{\text{actual}}$ applied to the set of such showers estimates the average atomic mass $<\ln{A_{\text{est}}}>$ of their shower initiating primaries, utilizing the $\eta_{\rho}$s. Results of the work revealed that the method of estimating the lateral shower age is almost high-energy interaction model-independent. A certain level of model dependence is albeit found in the case of $<\ln{A_{\text{est}}}>$ estimation.

\end{abstract}

\maketitle

\section{Introduction}
The origin and nature of cosmic rays (CRs) at energies $\gtrsim{1}$~PeV remain an enigma. A study of the mass composition of the CR flux is crucial to understanding these unresolved issues completely. The measured mass composition and also the energy spectrum of the CRs on Earth are known to be directly linked to the origin, acceleration, and propagation mechanisms of the highly energetic CR particles. However, the mass composition of CRs at these energies remains more poorly determined, and most of the results known to date from various experiments are still inconclusive.

Due to the steeply falling nature of its flux, the mass composition of CRs above energies of about 0.1 PeV is studied indirectly via extensive air shower (EAS) methods, examining the overall characteristics of EASs [1-2], and the longitudinal profile of the showers [3,4,5]. The current methods of detecting EASs are the measurements of secondary particle distributions at various observation levels and of deep underground muons, as well as the measurements of atmospheric fluorescence radiation, radio or Cherenkov emission originating from the EASs. Different characteristics namely the size of a shower (electron size or shower size and muon size), the shape of the lateral profiles (lateral shower age) and the shape of the longitudinal profiles (the depth of the shower maximum or the longitudinal shower age), essentially depend on the primary particle energy and the mass of the particle [6,7,8,9], can be estimated by arrays of particle detectors and fluorescence detectors on the ground. Furthermore, Cherenkov radiation detectors on the ground can only observe the lateral distribution of Cherenkov photons at the observation level, from which the height of the shower maximum, and then the depth of the shower maximum can be estimated [10]. On the other hand, the antennas of the radio antenna array [11] distributed over the ground, measure the lateral distribution of the radio emissions of CR EASs, which in turn provides information on the depth of the shower maximum and the radio amplitude, and later about the type and energy of the shower initiating primary particle [12].

The findings on the mass composition of CRs in the PeV region, even after significant experimental efforts, are still not enough to reach unanimity. The CR mass-sensitive EAS observables suffer greatly from the fluctuations of the EAS cascade sub-processes in air [13]. Moreover, the results depend on the equipment, the data analysis techniques, the properties of the hadronic interaction models, the atmospheric conditions, geomagnetic fields, etc. On the other hand, the efforts to measure the CR mass composition from the observations of Cherenkov [14] and fluorescence [14-15] radiations in the EeV regime are restricted due to uncertainties linked primarily with the model of hadronic interactions. Therefore, there is an attractive possibility to verify the previously obtained results on CR mass composition by a different method and a relatively new and simple EAS observable correlated to the shower age using the charged electromagnetic (EM) component of an EAS only. An alternative analysis method in this situation is likely to be more efficient, but we will not use it in this work. In this alternative approach, several existing and newly possible EAS observables (say, the present observable is one) are simultaneously used to deduce the mass composition of CRs [16-17].

The observed/simulated lateral density distribution (LDD) of electrons in a shower is consistently described by the well-known Nishimura-Kamata-Greisen (NKG) lateral density function (LDF) [18,19,20]. The lateral shower age parameter ($s$) in the LDF characteristically describes its slope [19,21], which is estimated routinely from the shape of the LDD electrons of a shower by applying shower reconstruction. On the contrary, the NKG-type LDF was a bit flatter than the observed LDD of electrons in EAS [22,23,24,25]. However, the LDD data at lower energies were better explained by the Monte Carlo (MC) methods of Hillas [26]. In a couple of our and other authors' works on the parameter $s$, we noticed that the LDF with a single $s$ is inadequate to accurately describe the simulated/observed LDD of electrons in an EAS at all distances [27,28,29,30,31]. All of these features suggest that the variable $r$ can characterize the parameter $s$ of an EAS, the radial distance from the EAS core. Such a radial dependency of $s$ was also found to be retained for EAS muons [31]. This feature of $s$ could not be eliminated even with some modified LDFs for the LDD of electrons/muons. At the outset, the idea of local shower age parameters (LAP) that emerged in [27-28] was justified. An MC calculation in [32] inferred that the mentioned radial dependency of $s$ is somewhat intermittent, resulting from an interplay of some likelihood effects during the EAS development and having little in common with the cascade theory. These effects on the LDFs may arise due to a more significant contribution of younger sub-cascades started by $\pi^{0}$s produced at the late stage of EAS development and also the finite transverse momenta of those $\pi^{0}$s. However, we substantiated that the shape of the LAP ($s_{local}$) versus radial distance curves (and hence the LDDs of the electrons/muons in an EAS) for a specific scale distance or the so-called Moli$\acute{e}$re radius ($r_m$) maintain nearly the same configuration i.e., their independence from the CR mass ($A$) and energy ($E$) [29,30,31,33]. All these crucial findings indicate that the LAP as a function of $r$ manifests some scaling nature, an intrinsic feature of a shower cascade.

Since $s$ of an EAS depends on $r$, its values obtained from various experiments will naturally be inconsistent, as the radial extension of the shower disks would differ from experiment to experiment. The same is even true for different EAS events in a single EAS experiment, as the radii of the shower disks fluctuate reasonably [34]. As the reliable and precise estimation of $s$ from the LDF fit to the measured electron densities or, in lieu, the inability of $s$ to describe the LDD of electrons in an EAS seems formidable, Capdevielle {\it et al.} suggested a complementary approach towards an unequivocal estimation of $s$. Thereupon, the notion of LAP and its various characteristics arose. The estimation of $s$ of an EAS from the variation of the LAP with radial distance ($r$) was started by Capdevielle {\it et al.} [27-28] and continued later by us [29,30,31]. One crucial feature of $s_{local}$ versus $r$ curve is that with an increase of $r$, the $s_{local}$ decreases initially and attains a minimum at about $40-50$~m, then it begins uprising, reaching a peak value, at about $300-320$~m, and then follows a descending trend again. This behaviour does not deviate considerably with the energy and mass of the CR particle [29] and also the shower size or muon size of the EAS [31]. The LAP emerges as an acceptable solution for exploring the nature associated with the lateral shower age of an EAS. Thus, the minimum LAP at the radial distance of about $45$~m [29] or an average of all LAPs falling in the radial distance range of $40-320$~m [30-31] was considered the lateral shower age (introduced in upcoming Sect. III by $s_{\text{av}}$) of an EAS.

The work reviewed so far concerning a more reliable and precise estimation of the shower age of an EAS, based on the two complementary approaches, requires a complete density distribution of electrons on the radial distance started off the EAS core. The universal nature of the configuration of the LDD of electrons in EAS, when described in terms of LAPs [29], offers a possibility to introduce a new, more basic observable, $\eta_{\rho(45;310)}$. The observable $\eta_{\rho(45;310)}$ is defined by the ratio of mean local electron densities centered at reference radial distances $45$~m ($\rho_{e;<r>=45m}$; corresponding to the minimum LAP of an EAS) and $310$~m ($\rho_{e;<r>=310m}$; corresponding to the maximum LAP of an EAS) from the EAS core. It was learned previously that in an extensively spread array ($<r>=350$~m), the local density of the EM component at a specific distance will be influenced weakly by the EAS cascade sub-processes in air [35-36]. Moreover, the systematic occurrence of the minimum LAP at $\sim 45$~m and the maximum LAP at $\sim 310$~m in an individual EAS is actually being substantiated here. The present work tried to use these two local densities of the EM component in the form of the new EAS observable $\eta_{\rho(45;310)}$ and to establish its correlation with the lateral shower age first (known as a mass-sensitive parameter in EAS studies) and then the mass number of the shower initiating CR particle. Our analysis of simulated showers also found that the local density of the charged EM component at $\sim 310$~m is very useful for converting it to the primary energy of the CR particle.

The plan of the paper is the following. The simple new observable $\eta_{\rho(45;310)}$ is formally introduced in section II. Section III describes the primary considerations of the MC simulations of air showers induced by protons and nuclei. A presentation of our results with the necessary discussion is given in section IV. Finally, conclusions are summarized in section V.

\section{A brief theoretical description of the observable}
Under some approximations (known as Approximation A \& B in the electromagnetic (EM) cascade theory) applied to cascade diffusion equations [37], the LDF of EAS particles progressing through the air was analytically expressed by Greisen and is well recognized as the NKG function, given by [18]
\begin{equation}
f(r)=C(s){(r/r_{m})}^{s-2}{(1+r/r_{m})}^{s-4.5}.
\end{equation}
Except for $C(s)$, called the normalization factor, all the remaining symbols in Eq.~(1) have already been introduced in Sect. I. Here, $C(s)$ is given by the following
\begin{equation}
C(s)=\frac{\Gamma{(4.5-s)}}{2\pi\Gamma{(s)}\Gamma{(4.5-2s)}}.
\end{equation}
Thanks to the properties of the Eulerian Beta function in the direct $r$-dependent part of the NKG function [37-38], a link can be established between one single cascade density, $\rho_{e}{(\text{r})}$ of electrons and $N_{e}$ via the profile function $f{(r)}$ as
\begin{equation}
\rho_{e}{(\mathrm{r})}=\frac{N_e}{r_{m}^{2}}f(r).
\end{equation}   
It should be noted that the above transformation exploited $s$ as a constant against $r$ for a specific atmospheric slant depth ($X$).\\

In Sect. I, we have explained in detail that $s$ of an EAS does depend on $r$, and Capdevielle et al. introduced the idea of local shower age, $s_{local}$ as a proposition [27-28]. The LAP emerged as a practical approach for examining the nature related to the lateral shower age of an EAS after its validation by the scientific world through important rapporteurs at international conferences [39,40,41].

\begin{figure}[!htbp]
	\centering
	\includegraphics[trim=0.6cm 0.6cm 0.6cm 0.6cm, scale=0.8]{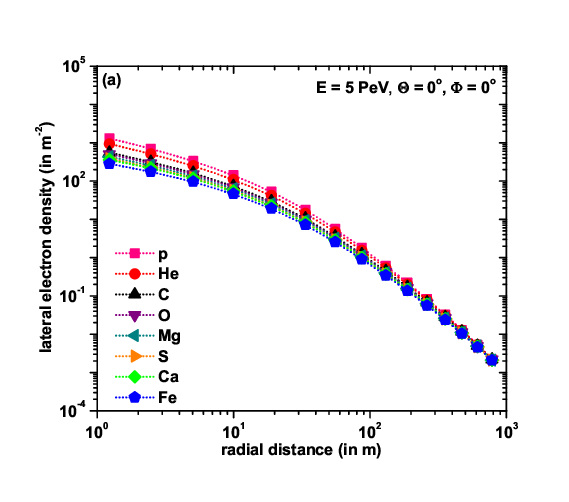}
	\includegraphics[trim=0.6cm 0.6cm 0.6cm 0.6cm, scale=0.8]{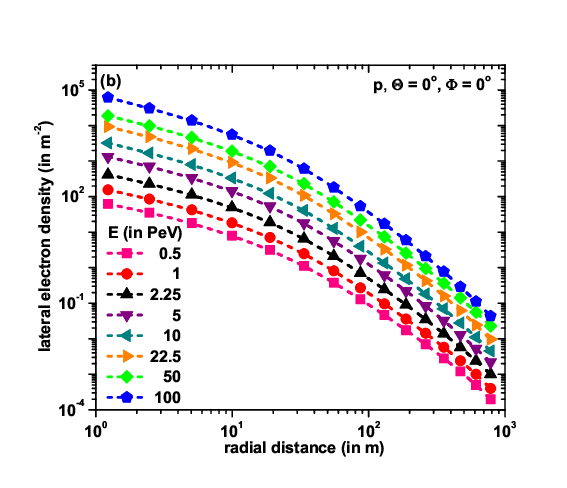}
	\caption{Lateral density distribution of shower electrons induced by (a) protons and nuclei for a fixed primary energy; (b) protons only for different fixed primary energies. The high-energy hadronic interaction model EPOS-LHC was embedded with the low-energy hadronic interaction model UrQMD in simulations. The lines are only a guide for the eye.}
\end{figure}

For the NKG-type LDF (say, $f(x)$ with $x=\frac{r}{r_m}$), by which lateral densities of electrons are described, the analytical expression of the local shower age for two adjacent points, $x_i$ and $x_j$ is given by     
\begin{equation}
s_{local}(i,j)=\frac{ln(F_{ij}X_{ij}^{2}Y_{ij}^{4.5})}{ln(X_{ij}Y_{ij})}. 
\end{equation}
Here, we use some substitutions like $F_{ij} = f(r_i)/f(r_j)$, $X_{ij} = r_i /r_j$ and $Y_{ij} = (x_i+1)/(x_j+1)$. The LAP at each point can be defined by making $r_{i}\rightarrow{r_{j}}$, and specifying $s_{local}(i,j)$ as $s_{local}(r)$ for $r=\frac{r_i+r_j}{2}$, and is given by
\begin{equation}
s_{\rm{local}}(r) = {1 \over {2x+1}} \left( (x+1) {{\partial{\ln f}} \over {\partial{\ln x}}} + 6.5x + 2 \right).
\end{equation}
This identification $s_{local}(r)=s_{local}(i,j)$ was found very decent for the experimental/simulated LDDs with $F_{ij}=\rho{(r_i)}/\rho{(r_j)}$ [29,31]. As the simulation for the work will be performed at the KASCADE level [42], $r_m$ will take $89$~m for the computation of the LAP.

\begin{figure}[!htbp]
	\centering
	\includegraphics[trim=0.6cm 0.6cm 0.6cm 0.6cm, scale=0.8]{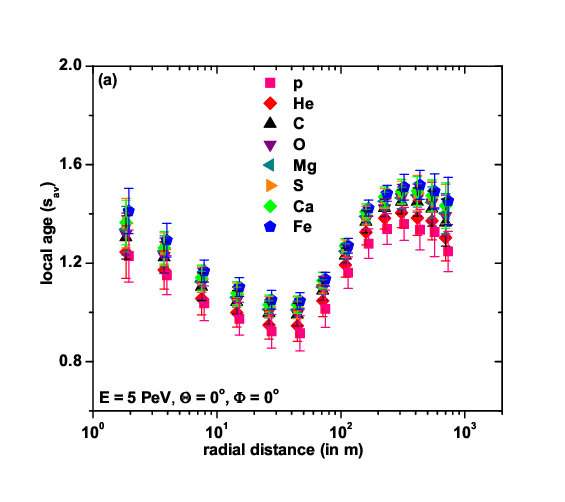}
	\includegraphics[trim=0.6cm 0.6cm 0.6cm 0.6cm, scale=0.8]{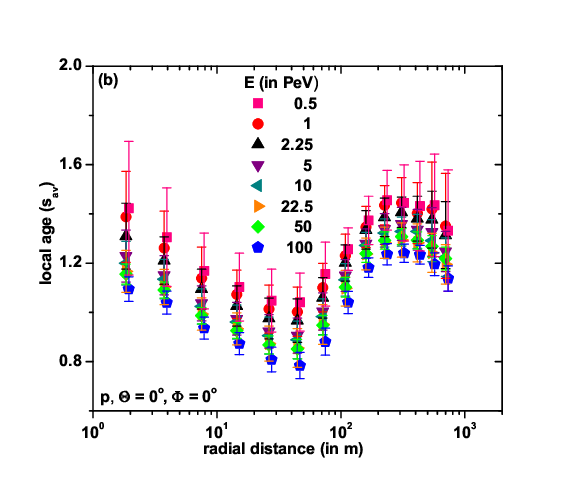}
	\caption{Variation of the local age parameter with radial distance for (a) different primaries with a fixed primary energy; (b) different primary energies with protons only. The Moliere radius $r_{m}$ is set at $89$~m. The high-energy hadronic interaction model EPOS-LHC was embedded with the low-energy hadronic interaction model UrQMD in simulations. We have shifted proton and Fe points (Fig.~a), and also the lowest and highest energy points (Fig.~b) slightly in the horizontal direction by multiplying each mean $r$ by $1.05$ to visually separate their errors from the rest.}
\end{figure}

\begin{table*}
\begin{center}
	\renewcommand{\arraystretch}{1.5}
	\begin{tabular}
	{| c | p{0.08\textwidth}  p{0.08\textwidth}   p{0.08\textwidth}   p{0.08\textwidth}   p{0.08\textwidth}  p{0.08\textwidth}   p{0.08\textwidth}   p{0.1\textwidth}  | } \hline
	\multicolumn{9}{|c|}{\bf Number of simulated showers [Group I (S-I or S-II sample)]} \\ \cline{1-9} 
 	\multirow{2}{*}{\rm{Type of PCR particle}} & \multicolumn{8}{c|}{Energy of PCR particle in PeV} \\ \cline{2-9} 
  	   		 &  0.5   &  1   & 2.25   & 5   & 10   & 22.5    & 50  & 100  \\ \hline
	 {\bf p}$^1_1$   &   50   &   50   &   50   &   50   &   50   &   50   &   30   &   10   \\ 
	{\bf He}$^4_2$   &   50   &   50   &   50   &   50   &   50   &   50   &   30   &   10   \\ 
	{\bf C}$^{12}_6$ &   50   &   50   &   50   &   50   &   50   &   50   &   30   &   10   \\ 
	{\bf O}$^{16}_8$ &   50   &   50   &   50   &   50   &   50   &   50   &   30   &   10   \\ 
{\bf Mg}$^{24}_{12}$ &   50   &   50   &   50   &   50   &   50   &   50   &   30   &   10   \\ 
 {\bf S}$^{32}_{16}$ &   50   &   50   &   50   &   50   &   50   &   50   &   30   &   10   \\ 
{\bf Ca}$^{40}_{20}$ &   50   &   50   &   50   &   50   &   50   &   50   &   30   &   10   \\ 			 
{\bf Fe}$^{56}_{26}$ &   50   &   50   &   50   &   50   &   50   &   50   &   30   &   10   \\ \hline   
	\multicolumn{8}{|l}{Total number of showers:} & 	\multicolumn{1}{l|}{2720} \\ \hline
	\end{tabular}
\caption {S-I sample contains 2720 number of simulated showers generated by EPOS-LHC hadronic interaction model. S-II sample also contains the same number of simulated showers but generated by QGSJet-II-04 hadronic interaction model.}
\end{center}
\end{table*}    

\begin{table*}
\begin{center}	
	\renewcommand{\arraystretch}{1.5}
	\begin{tabular}
	{| c | p{0.08\textwidth}  p{0.08\textwidth}   p{0.08\textwidth}   p{0.08\textwidth}  | } \hline
	\multicolumn{5}{|c|}{\bf Number of simulated showers [Group II (S-III or S-IV sample)]} \\ \cline{1-5} 
	\multirow{2}{*}{\rm{Type of PCR particle}} & \multicolumn{4}{c|}{Energy of PCR particle in PeV} \\ \cline{2-5} 
		&  2  &  8  & 32  & 128   \\ \hline
{\bf Li}$^{7}_{3}$ &   50   &   50   &   25   &   10    \\ 
{\bf N}$^{14}_{7}$ &   50   &   50   &   25   &   10    \\ 
{\bf Ne}$^{20}_{10}$ &   50   &   50   &   25   &   10    \\ 
{\bf Si}$^{28}_{14}$ &   50   &   50   &   25   &   10    \\ 
{\bf Ar}$^{40}_{18}$ &   50   &   50   &   25   &   10    \\ 
{\bf Cr}$^{52}_{24}$ &   50   &   50   &   25   &   10    \\ \hline      
 	\multicolumn{4}{|l}{Total number of showers:} & 	\multicolumn{1}{l|}{810} \\ \hline
	\end{tabular}
\caption {S-III sample contains 810 number of simulated showers generated by EPOS-LHC hadronic interaction model. S-IV sample also contains the same number of simulated showers but generated by QGSJet-II-04 hadronic interaction model.}
\end{center}
\end{table*} 

The LAP for the lateral densities of electrons was computed for each EAS event directly using Eq.~(4) [29,30,31]. The LAP versus $r$ curve has a distinctive feature where the parameter initially attains a minimum value at about $40-50$~m, then it shoots up, possessing a local maximum at about $300-320$~m, and then exhibits a falling trend again (see Fig.~2). This distinctive feature of the LAP concerning the LDD of electrons in EAS motivates us to revisit the concept of the lateral shower age of an EAS. The augmentation of the concept associated with the lateral shower age will be ascertained here by studying a new, more basic EAS observable, $\eta_{\rho(45;310)}$. The observable is defined by the ratio between the mean local electron densities computed at $45$~m and $310$~m radial distances from the EAS core as specified by the Fig.~1 in Sect.~III. $\eta_{\rho(45;310)}$ is then given by the simple relation.
\begin{equation}
\eta_{\rho(45;310)}=\frac{\rho_{e;<r>=45m}}{\rho_{e;<r>=310m}}. 
\end{equation}
To validate whether the new observable, $\eta_{\rho(45;310)}$ can achieve the objectives of the work indicated in Sect. I, we use an extensive analysis of simulated EAS events generated by using the air shower simulation package CORSIKA [43]. 

\section{Simulation of EASs and data analysis}
The simulation of the EAS events is performed using CORSIKA package version 7.7401 [43]. Two groups of primary nuclei are formed for simulation. Group I includes protons, helium, carbon, oxygen, magnesium, sulphur, calcium, and iron as primaries. On the other hand, the primary components of Group II are lithium, nitrogen, neon, silicon, argon and chromium. Two equal-sized samples (S-I \& S-II) of total $2\times{2720}$ EAS events are simulated using Group I nuclei as primaries corresponding to the high energy ($> 80~GeV/n$) hadronic interaction models EPOS-LHC [44] (S-I) and QGSJet-II-04 [45] (S-II). Two more samples (S-III \& S-IV) taking a minimal number, i.e. $2\times{810}$ of EAS events are also simulated using Group II nuclei, and EPOS-LHC (S-III) and QGSJet-II-04 (S-IV) as the high energy hadronic interaction models. This makes it possible to assess how well different empirical parameters found  through; (i) the study of the sample S-I may be applied to both the samples S-III and S-IV, and (ii) the study of the sample S-II may be applied to both the samples S-III and S-IV respectively. Both of these hadronic interaction models were used in combination with the low energy ($< 80~GeV/n$) hadronic interaction model UrQMD [46]. The EGS4 program library [47] takes care of the simulation of the EM component of an EAS in the EM sub-cascades of electrons and photons.\\
 
The simulated events are generated at the default geographical location of CORSIKA corresponding to the KASCADE experiment (latitude $49.1^o$~N, longitude $8.4^o$~E, 110~m a.s.l.) [42] at fixed zenith angle $\Theta = 0^{\circ}$ only. The kinetic energy thresholds for muons and electrons are set at $0.3$ and $0.003$~GeV. Tables I and II present details about four simulated samples, S-I, S-II, S-III and S-IV, of EAS events with different composition series, primary energies, and hadronic interaction models.\\
\begin{figure*}[!htbp]
	\subfigure
	{\includegraphics[trim=0.6cm 0.6cm 0.0cm 0.6cm, scale=0.8]{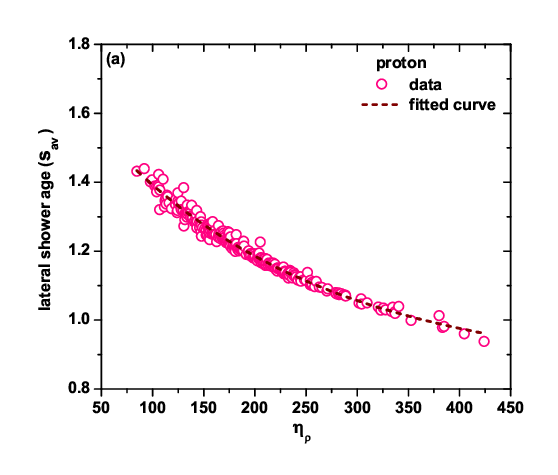}}
	\subfigure
	{\includegraphics[trim=0.6cm 0.6cm 0.0cm 0.6cm, scale=0.8]{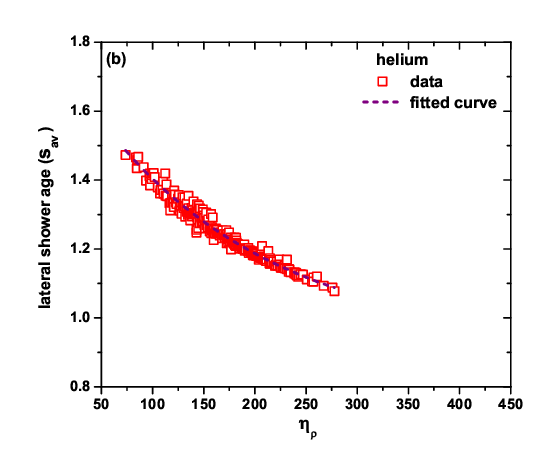}}
	\\      \subfigure 
	{\includegraphics[trim=0.6cm 0.6cm 0.0cm 0.6cm, scale=0.8]{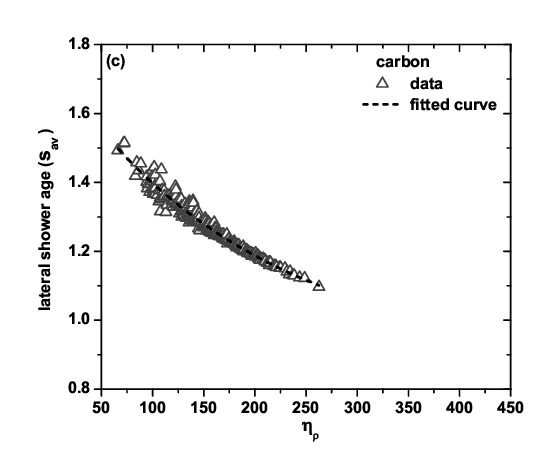}} 
	\subfigure
	{\includegraphics[trim=0.6cm 0.6cm 0.6cm 0.6cm, scale=0.8]{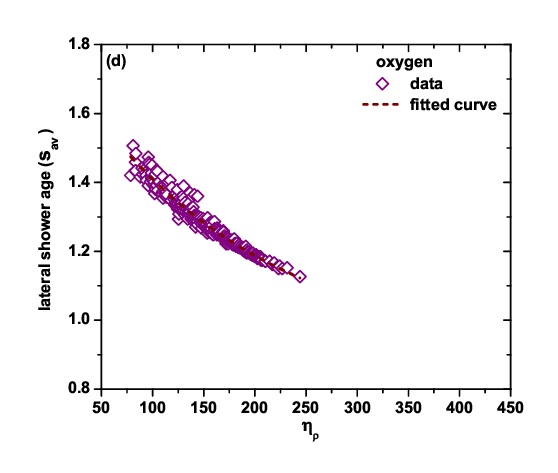}}
	\\      \subfigure
	{\includegraphics[trim=0.6cm 0.6cm 0.0cm 0.6cm, scale=0.8]{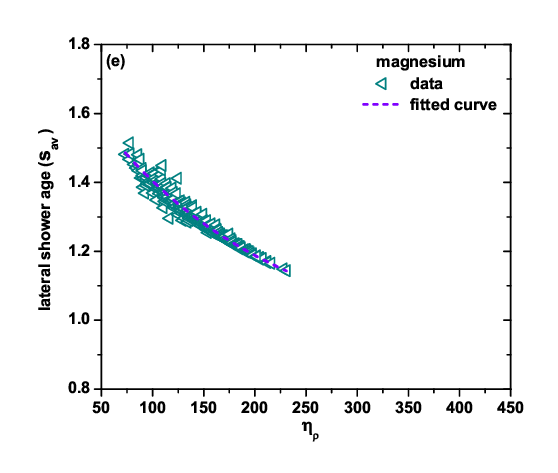}}
	\subfigure 
	{\includegraphics[trim=0.6cm 0.6cm 0.6cm 0.6cm, scale=0.8]{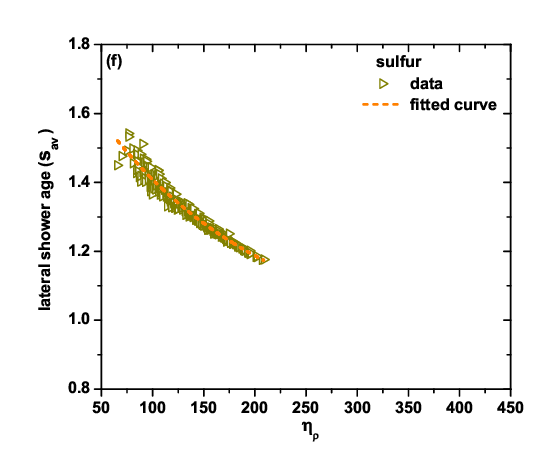}}
	\\      \subfigure 
	{\includegraphics[trim=0.6cm 0.6cm 0.0cm 0.6cm, scale=0.8]{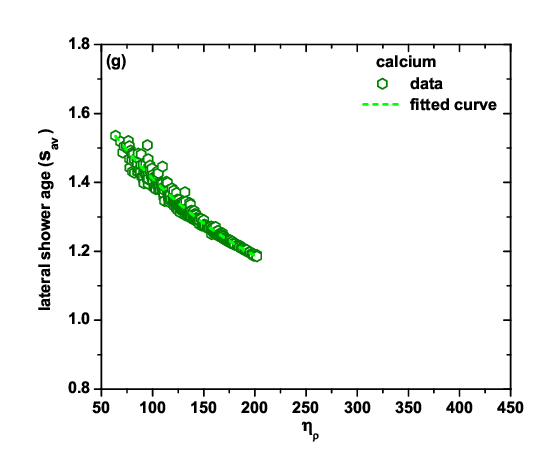}} 
	\subfigure
	{\includegraphics[trim=0.6cm 0.6cm 0.6cm 0.6cm, scale=0.8]{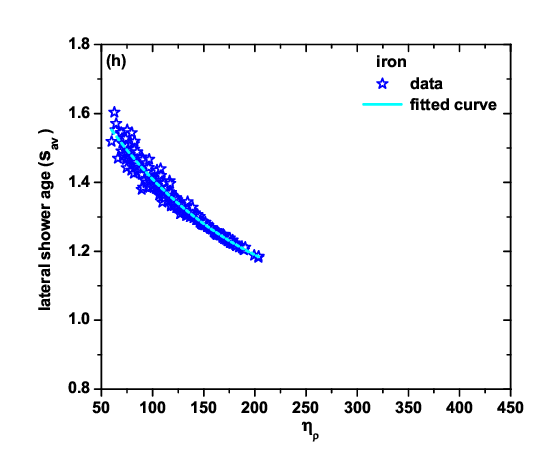}}	
	\caption{Variation of the lateral shower age parameter $s_{\text{av}}$ with $\eta_{\rho}$ for different showers induced by Group I primaries using the EPOS-LHC model (S-I). Each type of shower has been generated at eight different primary energies.}
\end{figure*}

The LAP of a simulated/observed CR shower has been estimated using the main Eq.~(4). Such an equation requires only two basic information on the density and distance data from the LDD of electrons of a shower [29,31]. For estimating the basic density data, we have subdivided the entire radial distance range $1-1000$~m in an EAS into $14$ fixed distance bands ($\Delta$) on the log scale by setting, ${\Delta}_{ji}=\ln{r}_{j}-\ln{r}_{i}\approx{0.5}$ with increasing $r$. In practice, however, this procedure may not be suitable at any arbitrary radial distance owing to considerable uncertainties in density and EAS core measurements and also from array triggering conditions. However, some of these conditions required by the present method could be practicable in EAS experiments utilizing closely packed particle detectors. Moreover, a reasonable number of showers must be analyzed to reduce the statistical fluctuations over electron densities at various radial distance intervals. 

In Fig.1a, we have studied the variation of the simulated electron densities over the range $1-1000$~m at a fixed $E$ with the EPOS-LHC model for eight different primary nuclei taken for the sample  S-I. For a particular primary species $p$, the nature of variation of the simulated electron densities with radial distance is shown in Fig.1b for $8$ different fixed primary energies. The variation of the LAP with radial distance from the EAS core is the main attribute from which the concept of the observable $\eta_{\rho(45;310)}$ has emerged. A study of this kind is depicted in Fig.~2 per the electron density data used in Fig.~1a (here Fig.~2a) and b (here Fig.~2b). The LAP estimation is affected mixedly by the radial distance range, primary energy, and CR composition. It is found that errors in LAP estimation are more significant for lighter nuclei-initiated showers than those initiated by heavier nuclei. 

It is important to note that the core position for a simulated shower is always situated at the origin ($0,0$) in the intrinsic 2D CORSIKA shower plane [43]. Each secondary particle in an EAS reaching the ground CORSIKA plane corresponding to $z=0$ has their ($x,y$) coordinates, and in terms of these coordinates the radial distance $r$ of the particle from the EAS core can be determined. In simulation, the radial distance of each of the EAS particles is known with great precision. However, the dispersion of the LAPs of an average EAS lies in the range $\sim{0.05-0.15}$ for $10~m<r<750~m$. The two average distances $45$~m and $310$~m were chosen from the characteristic high-low-high kind of radial variation in the LAP within $\sim 330$~m or so (Fig. 2). The figure shows that with an increase of $r$, the LAP experiences a gradual decrement, reaching a least possible value at about $45$~m, and starts increasing, and attained a maximum at a radial distance of about $310$~m. The average lateral electron densities at these two radial distances ($45$ and $310$~m) were used to have the new parameter $\eta_{\rho}$.

\begin{figure}[!htbp]
	\centering
	\includegraphics[trim=0.6cm 0.6cm 0.6cm 0.6cm, scale=0.8]{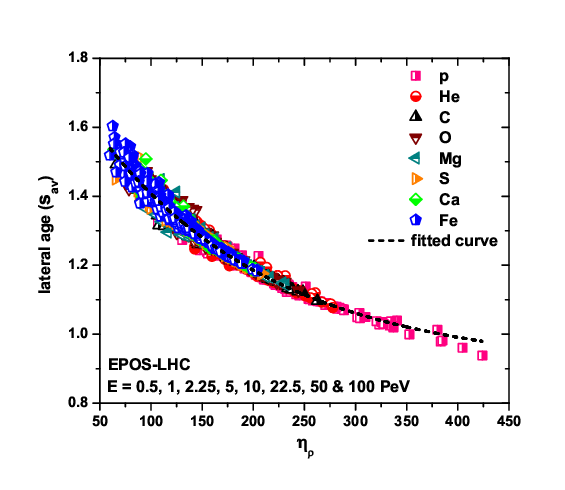}
	\caption{Variation of the lateral shower age parameter $s_{\text{av}}$ with $\eta_{\rho}$ for all 2720 showers of Group I (S-I), generated at eight different primary energies.}
\end{figure}

\begin{table*}
	\begin{center}
		\begin{tabular}{p{0.18 \textwidth} p{0.2\textwidth}  p{0.2\textwidth}   p{0.2\textwidth}  }
			\cline{1-4}
			PCR particle& {\bf $a$} 			& {\bf $b$} 			& {\bf $c$}  					\\ \hline
			p   			& 0.8426 $\pm$ 0.0149 & 0.8810 $\pm$ 0.0075	& 212.2132 $\pm$ 9.2473	   	\\ 
			He   			& 0.9184 $\pm$ 0.0206 & 0.8777 $\pm$ 0.0082	& 168.7812 $\pm$ 10.4938   	\\ 	
			C   			& 0.8735 $\pm$ 0.0198 & 0.8769 $\pm$ 0.0198	& 194.7877 $\pm$ 17.0395  	\\ 
			O   			& 0.8843 $\pm$ 0.0430 & 0.9077 $\pm$ 0.0218	& 182.4862 $\pm$ 19.4597  	\\ 	
			Mg   			& 0.9126 $\pm$ 0.0414 & 0.8759 $\pm$ 0.0209	& 172.9396 $\pm$ 18.4137  	\\ 
			S   			& 0.9256 $\pm$ 0.0433 & 0.8902 $\pm$ 0.0213	& 163.4759 $\pm$ 18.0517  	\\ 	
			Ca   			& 0.9316 $\pm$ 0.0349 & 0.9014 $\pm$ 0.0175 & 158.4254 $\pm$ 13.9781  	\\ 
			Fe   			& 0.9780 $\pm$ 0.0328 & 0.8810 $\pm$ 0.0146 & 140.1258 $\pm$ 12.5557   	\\ \hline 
			From Fig.~4 only   & $\bf{0.9013\pm{0.0069}}$
			&  $\bf{0.9007\pm{0.0036}}$ 
			&  $\bf{173.5525\pm{3.1992}}$	   	\\ \hline								
		\end{tabular}
\caption{Values for parameters $a$, $b$ and $c$ that were determined from the fitting of the $s_{\text{av}}$ versus $\eta_{\rho}$ data presented in Fig.~3 and Fig.~4 by the Eq.~(7). The high-energy hadronic interaction model EPOS-LHC (S-I) was embedded with the low-energy hadronic interaction model UrQMD in simulations. The highlighted numerals in the last row show the combined fit parameters $a,b,c$ from  Fig.~4 only.}
	\end{center}
\end{table*}

\begin{figure*}[!htbp]
	\subfigure
	{\includegraphics[trim=0.6cm 0.6cm 0.0cm 0.6cm, scale=0.8]{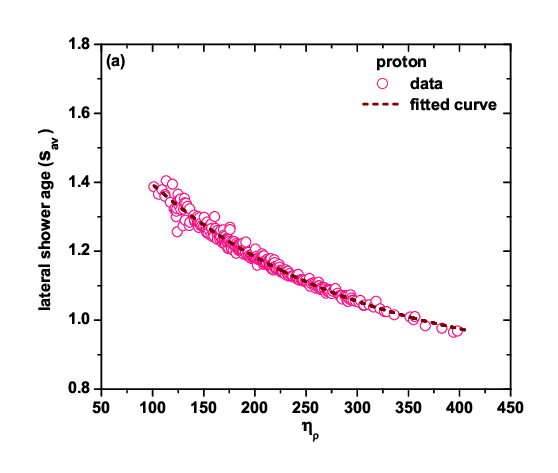}}
	\subfigure
	{\includegraphics[trim=0.6cm 0.6cm 0.0cm 0.6cm, scale=0.8]{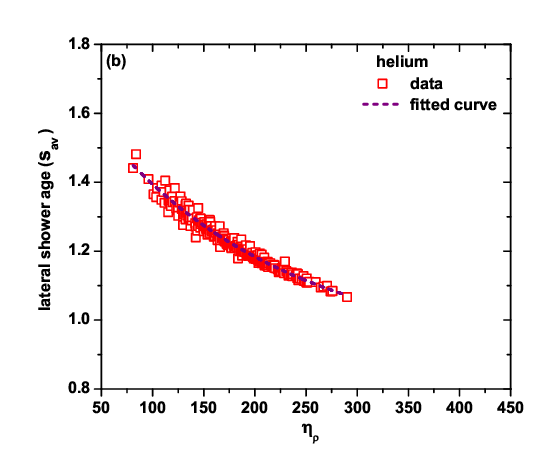}}
	\\      \subfigure 
	{\includegraphics[trim=0.6cm 0.6cm 0.0cm 0.6cm, scale=0.8]{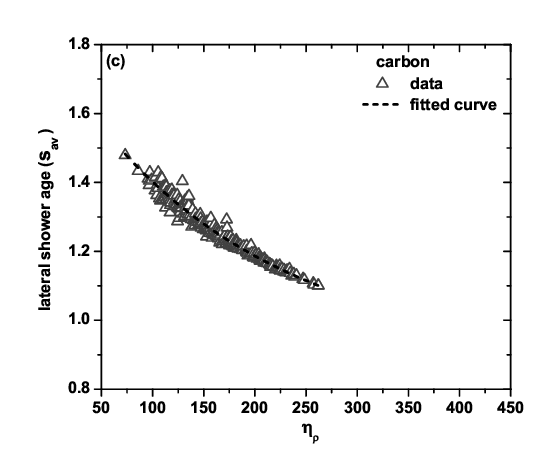}} 
	\subfigure
	{\includegraphics[trim=0.6cm 0.6cm 0.6cm 0.6cm, scale=0.8]{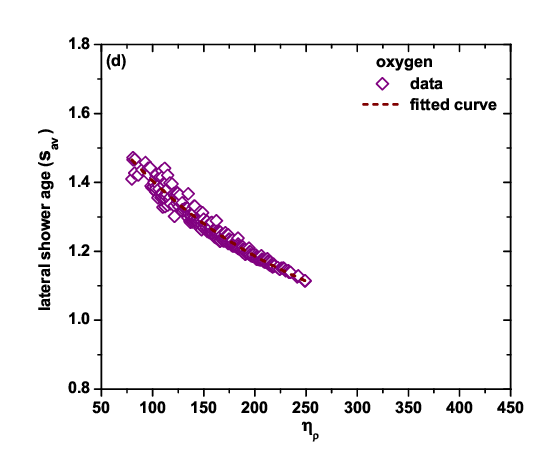}}
	\\      \subfigure
	{\includegraphics[trim=0.6cm 0.6cm 0.0cm 0.6cm, scale=0.8]{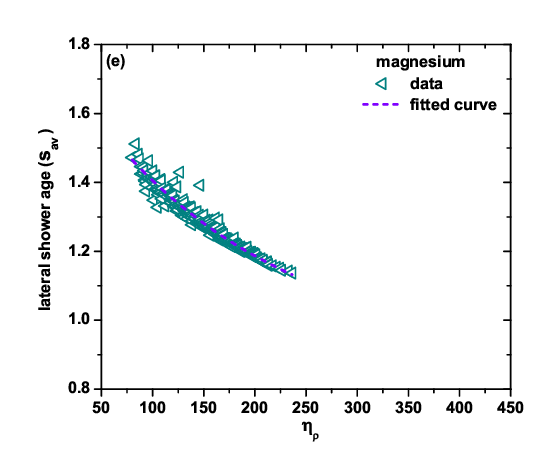}}
	\subfigure 
	{\includegraphics[trim=0.6cm 0.6cm 0.6cm 0.6cm, scale=0.8]{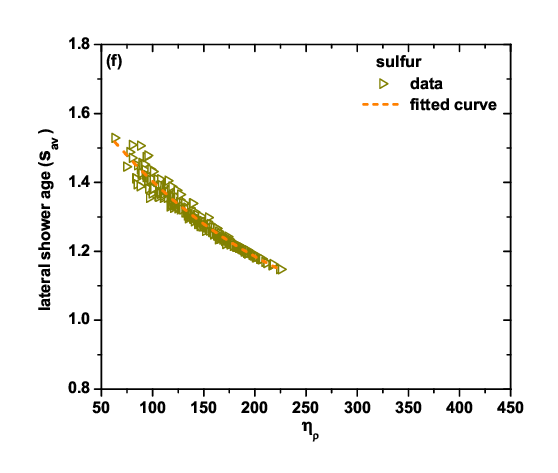}}
	\\      \subfigure 
	{\includegraphics[trim=0.6cm 0.6cm 0.0cm 0.6cm, scale=0.8]{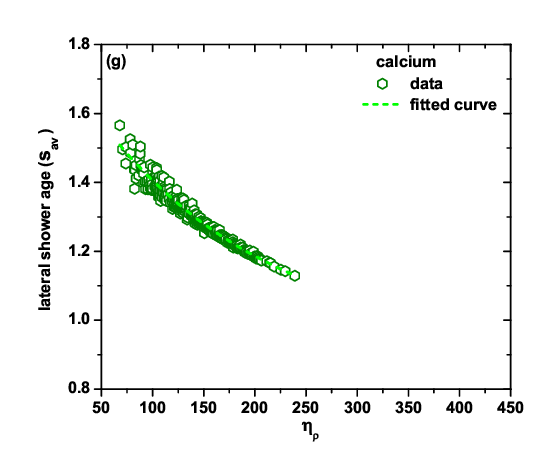}} 
	\subfigure
	{\includegraphics[trim=0.6cm 0.6cm 0.6cm 0.6cm, scale=0.8]{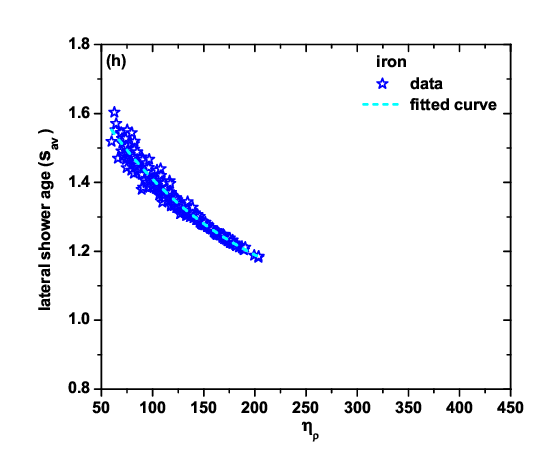}}	
	\caption{Variation of the lateral shower age parameter $s_{\text{av}}$ with $\eta_{\rho}$ for different showers induced by Group I (S-II) primaries using the QGSJet-II-04 model. Each type of shower has been generated at eight different primary energies.}
\end{figure*}

\begin{figure}[!htbp]
	\centering
	\includegraphics[trim=0.6cm 0.6cm 0.6cm 0.6cm, scale=0.8]{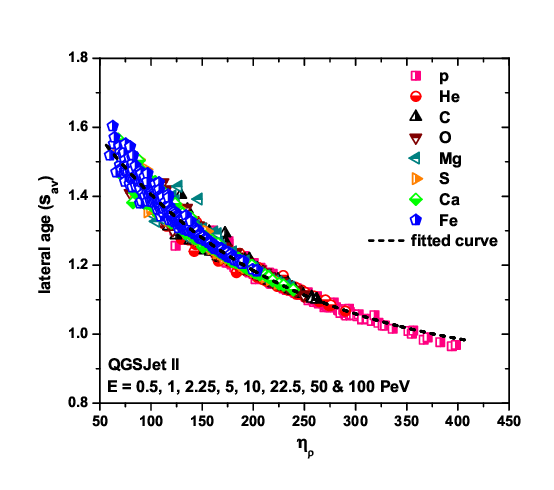}
	\caption{Variation of the lateral shower age parameter $s_{\text{av}}$ with $\eta_{\rho}$ for all 2720 showers in Group I (S-II), generated at eight different primary energies using QGSJet-II-04 model.}
\end{figure}

\begin{table*}
	\begin{center}
		\begin{tabular}{p{0.18 \textwidth} p{0.2\textwidth}  p{0.2\textwidth}   p{0.2\textwidth}  }
			\cline{1-4}
			PCR particle& {\bf $a$} 			& {\bf $b$} 			& {\bf $c$} 				\\ \hline
			p   			& 0.8495 $\pm$ 0.0201 & 0.8829 $\pm$ 0.0077	& 206.2607 $\pm$ 12.4681   	\\ 
			He   			& 0.8957 $\pm$ 0.0259 & 0.8626 $\pm$ 0.0089	& 182.3864 $\pm$ 12.9568   	\\ 	
			C   			& 0.8872 $\pm$ 0.0330 & 0.8854 $\pm$ 0.0140	& 184.3970 $\pm$ 16.6556  	\\ 
			O   			& 0.8824 $\pm$ 0.0359 & 0.8985 $\pm$ 0.0173	& 184.3331 $\pm$ 17.0848  	\\ 	
			Mg   			& 0.8869 $\pm$ 0.0441 & 0.8967 $\pm$ 0.0222	& 182.4826 $\pm$ 20.2601  	\\ 
			S   			& 0.8886 $\pm$ 0.0404 & 0.8926 $\pm$ 0.0088	& 181.5454 $\pm$ 18.9039  	\\ 	
			Ca   			& 0.9151 $\pm$ 0.0357 & 0.8881 $\pm$ 0.0175 & 168.5360 $\pm$ 15.6733  	\\ 
			Fe   			& 0.9780 $\pm$ 0.0329 & 0.8809 $\pm$ 0.0148 & 140.1266 $\pm$ 12.6263   	\\ \hline 
			From Fig.~6 only   & $\bf{0.8922\pm{0.0075}}$
			&  $\bf{0.8985\pm{0.0036}}$ 
			&  $\bf{178.3235\pm{3.6313}}$	   	\\ \hline								
		\end{tabular}
\caption{Values for parameters $a$, $b$ and $c$ that were determined from the fitting of the $s_{\text{av}}$ versus $\eta_{\rho}$ data presented in Fig.~5 and Fig.~6 by the Eq.~(7). The high-energy hadronic interaction model QGSJet-II-04 (S-II) was embedded with the low-energy hadronic interaction model UrQMD in simulations. The highlighted numerals in the last row show the combined fit parameters $a,b,c$ from  Fig.~6 only.}
	\end{center}
\end{table*}

In this work, we have seen that the event-by-event analysis of simulated (or even in observed data [29,31]) electron density data provides a number (here, $14$) of LAPs for each shower from the specified $14$ radial distance bands. One observable parameter was then picked to look into any possible physical character related to the LAP so that it is sensitive enough to the mass composition of the CR primary [29]. Such an observable extracted from the LAP can be obtained by taking an average value ($s_{\text{av}}$) of several LAPs in the radial distance range $45-310$~m. It is noteworthy to mention that such an observable $s_{\text{av}}$ was considered (a lateral shower age obtained from the LAP) and used in our earlier works [29,30,31].

Here, we will demonstrate a strong correlation between $\eta_{\rho(45;310)}$ (henceforth $\eta_{\rho}$ in short) and $s_{\text{av}}$ from the analysis of simulated showers induced by nuclei in Group I. The empirical relation describing such a correlation will then be applied to the simulated showers induced by nuclei in Group II to estimate the lateral shower age (we denote it by $s_{estimated}$, and henceforth by $s_{est}$ in short) using only the key observable $\eta_{\rho}$ for each shower in S-III and S-IV samples. The paper divulges a new approach in which $\eta_{\rho}$ may act as a primary observable to characterize CR showers and estimate the mass $A$ of the EAS initiating primary.
\begin{table*}
\begin{center}
	\renewcommand{\arraystretch}{1.2}
	\begin{tabular}{p{0.14 \textwidth} p{0.20\textwidth}  p{0.2\textwidth}   p{0.15\textwidth}   p{0.15\textwidth} p{0.08\textwidth} } \hline
	PCR particle  & PCR energy [in PeV] & $\eta_{\rho}\pm{\Delta{\eta_{\rho}}}$  & $s_{av}\pm{\Delta{s_{av}}}$ & $s_{est}\pm{\Delta{s_{est}}}$ \\\hline
\multirow{4}{*}{\rm{Li}} 	&  $0.2$  &  151.443 $\pm$ 21.515 & 1.279 $\pm$ 0.044 & 1.277 $\pm$ 0.047	\\ 
							&  $8$  &  174.433 $\pm$ 19.083 & 1.232 $\pm$ 0.034 & 1.231 $\pm$ 0.037	\\
							& $32$  &  209.378 $\pm$ 20.506	& 1.174 $\pm$ 0.030 & 1.170 $\pm$ 0.033	\\
							&$128$  &  240.146 $\pm$ 13.266 & 1.128 $\pm$ 0.016 & 1.127 $\pm$ 0.019	\\	\hline						
\multirow{4}{*}{\rm{N}} 	&  $0.2$  &  136.221 $\pm$ 18.496 & 1.311 $\pm$ 0.038 & 1.312 $\pm$ 0.044	\\ 
							&  $8$  &  162.554 $\pm$ 17.209 & 1.255 $\pm$ 0.031 & 1.254 $\pm$ 0.0362	\\
							& $32$  &  187.139 $\pm$ 15.485 & 1.207 $\pm$ 0.025 & 1.207 $\pm$ 0.028	\\
							&$128$  &  222.224 $\pm$ 17.999 & 1.153 $\pm$ 0.026 & 1.151 $\pm$ 0.027	\\	\hline	
\multirow{4}{*}{\rm{Ne}} 	&  $0.2$  &  130.470 $\pm$ 17.485 & 1.326 $\pm$ 0.037 & 1.326 $\pm$ 0.043	\\ 
							&  $8$  &  156.085 $\pm$	14.738 & 1.268 $\pm$ 0.028 & 1.267 $\pm$ 0.032	\\
							& $32$  &  186.973 $\pm$  9.444 & 1.209 $\pm$ 0.016 & 1.208 $\pm$ 0.019	\\
							&$128$  &  212.229 $\pm$	 8.163 & 1.168 $\pm$ 0.011 & 1.166 $\pm$ 0.015	\\	\hline	
\multirow{4}{*}{\rm{Si}} 	&  $0.2$  &  126.406 $\pm$ 14.135 & 1.336 $\pm$ 0.036 & 1.336 $\pm$ 0.036	\\ 
							&  $8$  &  153.176 $\pm$ 14.092 & 1.274 $\pm$ 0.029 & 1.273 $\pm$ 0.031	\\
							& $32$  &  179.557 $\pm$	10.852 & 1.222 $\pm$ 0.019 & 1.221 $\pm$ 0.022	\\
							&$128$  &  203.765 $\pm$	11.273 & 1.181 $\pm$ 0.016 & 1.179 $\pm$ 0.020	\\	\hline						
\multirow{4}{*}{\rm{Ar}} 	&  $0.2$  &  118.260 $\pm$ 11.288 & 1.357 $\pm$ 0.027 & 1.357 $\pm$ 0.031 \\ 
							&  $8$  &  145.607 $\pm$ 11.837	& 1.290 $\pm$ 0.024 & 1.290 $\pm$ 0.028	\\
							& $32$  &  169.287 $\pm$ 12.898 & 1.241 $\pm$ 0.022 & 1.240 $\pm$ 0.026	\\
							&$128$  &  200.623 $\pm$	 9.450 & 1.186 $\pm$ 0.013 & 1.184 $\pm$ 0.018	\\	\hline	
\multirow{4}{*}{\rm{Cr}} 	&  $0.2$  &  113.585 $\pm$ 10.208 & 1.369 $\pm$ 0.027 & 1.369 $\pm$ 0.029	\\ 
							&  $8$  &  139.746 $\pm$  9.325 & 1.303 $\pm$ 0.020 & 1.303 $\pm$ 0.023	\\
							& $32$  &  167.917 $\pm$ 11.270 & 1.245 $\pm$ 0.019 & 1.243 $\pm$ 0.024	\\
							&$128$  &  189.641 $\pm$	11.383 & 1.203 $\pm$ 0.018 & 1.203 $\pm$ 0.021	\\	\hline	
		\end{tabular}
\caption {The average values of $\eta_{\rho}$, $s_{\text{av}}$ and $s_{\text{est}}$ for all the $810$ simulated showers in S-III (EPOS-LHC) sample of Group II.} 
	\end{center}
\end{table*}  

\begin{figure}[!htbp]
	\centering
	\includegraphics[trim=0.6cm 0.6cm 0.6cm 0.6cm, scale=0.8]{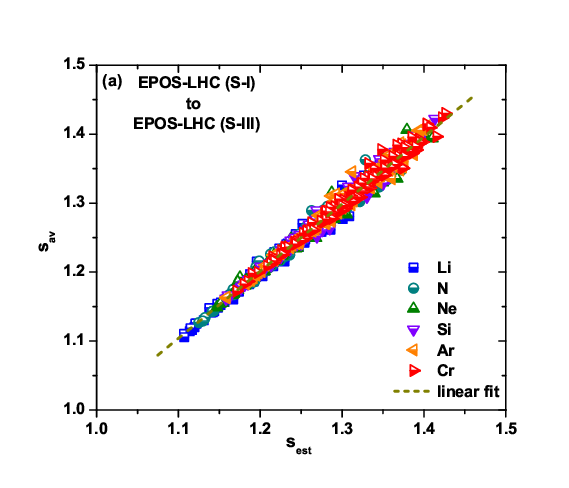}
	\includegraphics[trim=0.6cm 0.6cm 0.6cm 0.6cm, scale=0.8]{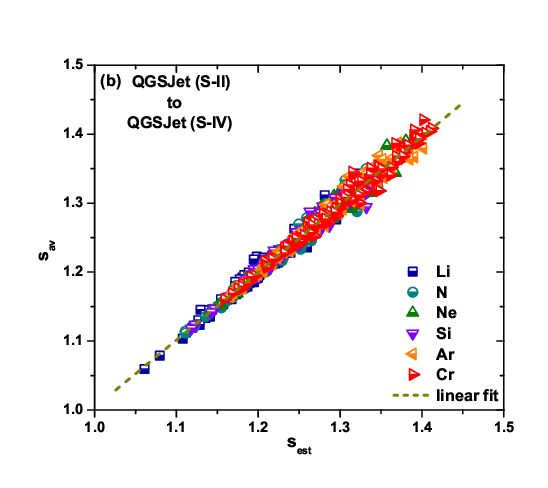}
  \includegraphics[trim=0.6cm 0.6cm 0.6cm 0.6cm, scale=0.8]{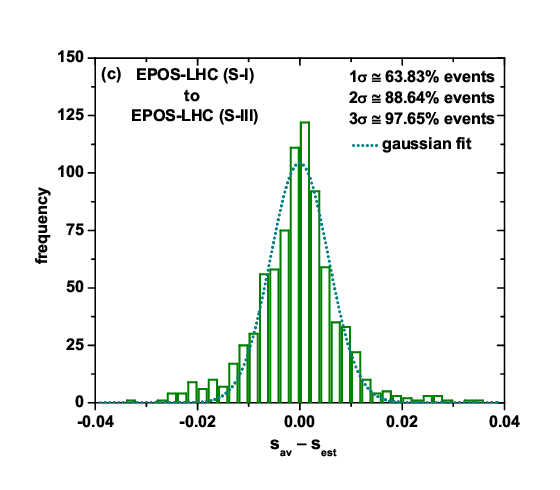}
	\includegraphics[trim=0.6cm 0.6cm 0.6cm 0.6cm, scale=0.8]{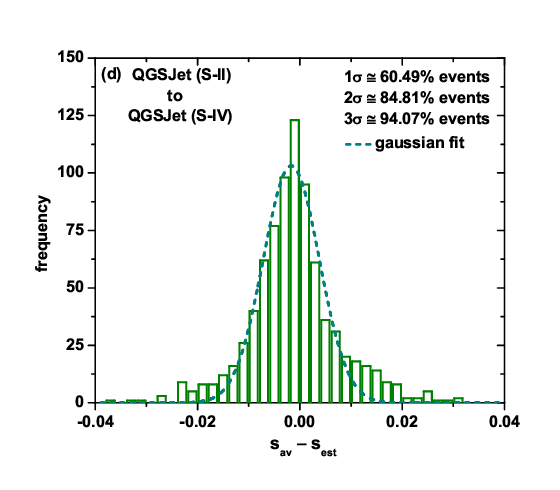}
	\caption{Fig.~a: the lateral shower age parameters $s_{\text{av}}$ obtained from the averaging of several LAPs in the radial distance range $45-310$~m as a function of the estimated lateral age parameters $s_{\text{est}}$ for S-III showers using the combined fit parameters $a,b,c$ obtained from the analysis of S-I showers; Fig.~b: Same as Fig.~a but between S-II and S-IV samples. The correlation between $s_{\text{av}}$ and $s_{\text{est}}$ has been approximated by $s_{\text{av}}=a_{1}~{s_{\text{est}}}+s_{0}$; Fig.~c \& d: distributions of $s_{\text{av}}-s_{\text{est}}$ for S-III and S-IV showers respectively.}
\end{figure}

\begin{figure}[!htbp]
	\centering
	\includegraphics[trim=0.6cm 0.6cm 0.6cm 0.6cm, scale=0.8]{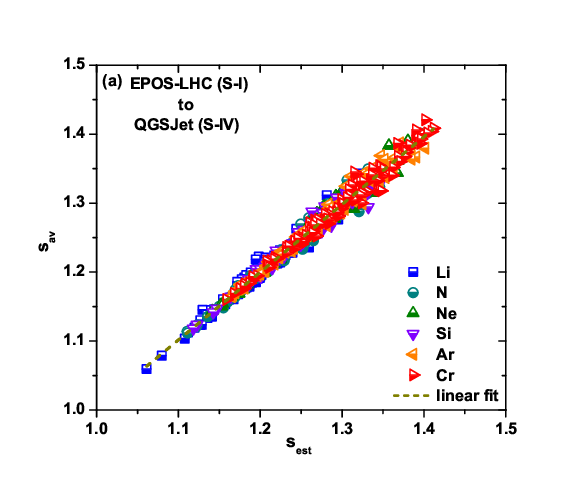}
	\includegraphics[trim=0.6cm 0.6cm 0.6cm 0.6cm, scale=0.8]{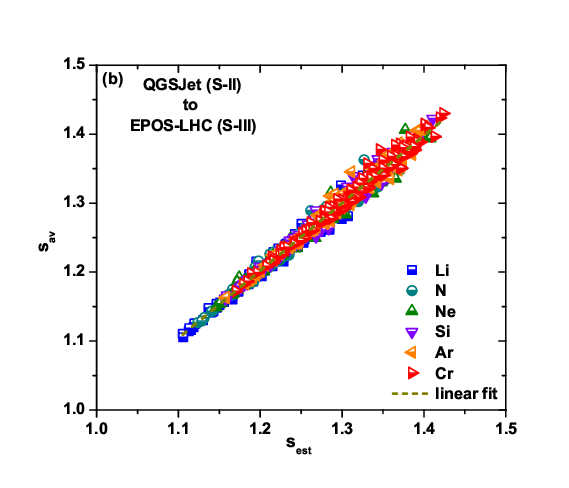}
  \includegraphics[trim=0.6cm 0.6cm 0.6cm 0.6cm, scale=0.8]{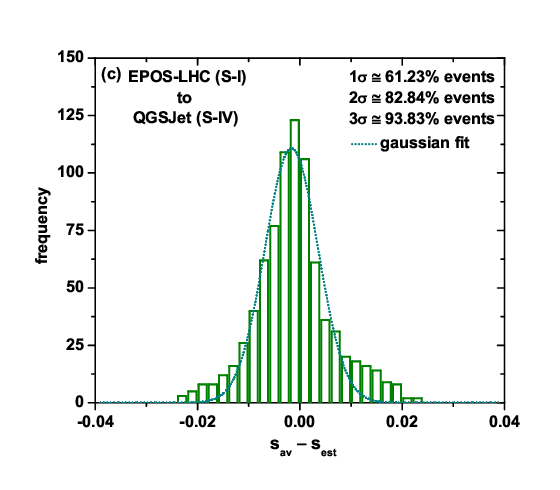}
	\includegraphics[trim=0.6cm 0.6cm 0.6cm 0.6cm, scale=0.8]{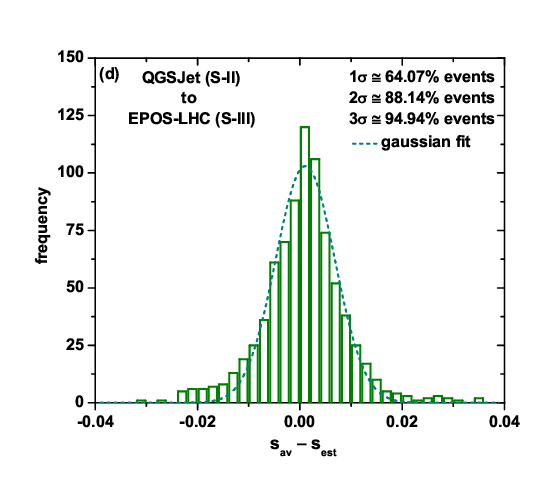}
	\caption{Fig.~a: the lateral shower age parameters $s_{\text{av}}$ obtained from the averaging of several LAPs in the radial distance range $45-310$~m as a function of the estimated lateral age parameters $s_{\text{est}}$ for S-IV showers using the combined fit parameters $a,b,c$ obtained from the analysis of S-I showers; Fig.~b: Same as Fig.~a but between S-II and S-III samples. The correlation between $s_{\text{av}}$ and $s_{\text{est}}$ has been approximated by $s_{\text{av}}=a_{1}~{s_{\text{est}}}+s_{0}$; Fig.~c \& d: distributions of $s_{\text{av}}-s_{\text{est}}$ for S-IV and S-III showers respectively.}
\end{figure}

\begin{figure}[!htbp]
	\centering
	\includegraphics[trim=0.6cm 0.6cm 0.6cm 0.6cm, scale=0.8]{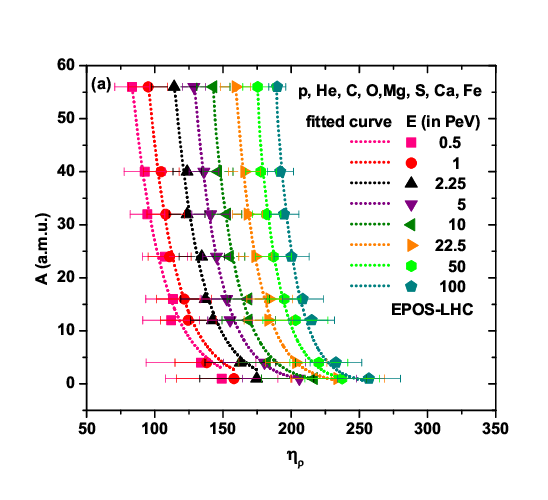}
	\includegraphics[trim=0.6cm 0.6cm 0.6cm 0.6cm, scale=0.8]{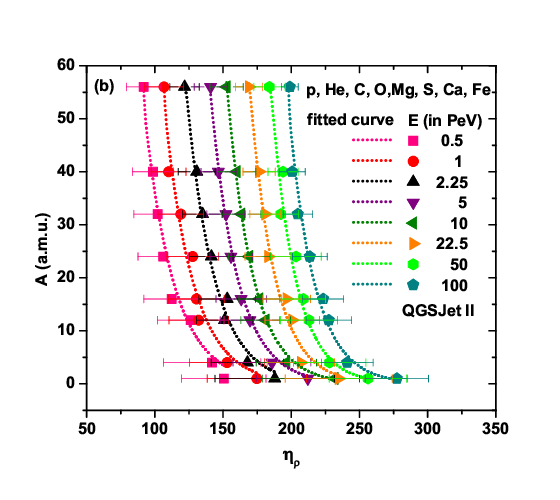}
	\caption{Fig.~a: variation of $\mathrm{A}$ with $\eta_{\rho}$ for S-I nuclei corresponding to eight different primary energies (EPOS-LHC model). Fig.~b: same as Fig.~a but for the S-II sample (QGSJet-II-04 model).}
\end{figure}

\begin{table*}[!htbp]
	\begin{center}
		\begin{tabular}{|p{0.12 \textwidth} p{0.16\textwidth}  p{0.16\textwidth} | }
			\cline{1-3}
			E (PeV)& {\bf $\alpha(E)$} 			& {\bf $\beta(E)$} \\ \hline
			0.5   			& -0.0447 $\pm$ 0.0029 & ~7.7713 $\pm$ 0.3965	\\ 
			1   			& -0.0484 $\pm$ 0.0028 & ~8.6508 $\pm$ 0.4451  	\\ 	
			2.25   			& -0.0505 $\pm$ 0.0028 & ~9.8139 $\pm$ 0.4846	\\ 
			5   			& -0.0531 $\pm$ 0.0021 & 10.8741 $\pm$ 0.2319 	\\ 	
			10   			& -0.0556 $\pm$ 0.0025 & 11.9164 $\pm$ 0.3093 	\\ 
			22.5   			& -0.0589 $\pm$ 0.0027 & 13.4106 $\pm$ 0.2824 	\\ 	
			50   			& -0.0631 $\pm$ 0.0024 & 15.0286 $\pm$ 0.2662  	\\ 
			100   			& -0.0649 $\pm$ 0.0024 & 16.0235 $\pm$ 0.2598  	\\ \hline 
		\end{tabular}
\caption {The parameters $\alpha$ and  $\beta$ at different primary energies for showers initiated by S-I nuclei (EPOS-LHC model).}
	\end{center}
\end{table*}

\begin{table*}[!htbp]
	\begin{center}
		\begin{tabular}{|p{0.12 \textwidth} p{0.16\textwidth}  p{0.16\textwidth} | }
				\cline{1-3}
				E (PeV)& {\bf $\alpha(E)$} 			& {\bf $\beta(E)$} \\ \hline
				0.5   			& -0.0463 $\pm$ 0.0029 & ~8.2194 $\pm$ 0.4865	\\ 
				1   			& -0.0477 $\pm$ 0.0028 & ~9.0595 $\pm$ 0.4463  	\\ 	
				2.25   			& -0.0492 $\pm$ 0.0029 & 10.0620 $\pm$ 0.3913	\\ 
				5   			& -0.0509 $\pm$ 0.0025 & 11.1704 $\pm$ 0.3153 	\\ 	
				10   			& -0.0529 $\pm$ 0.0025 & 12.1158 $\pm$ 0.3089 	\\ 
				22.5   			& -0.0549 $\pm$ 0.0027 & 13.3384 $\pm$ 0.2819 	\\ 	
				50   			& -0.0567 $\pm$ 0.0022 & 14.5185 $\pm$ 0.2671  	\\ 
				100   			& -0.0586 $\pm$ 0.0022 & 15.5862 $\pm$ 0.2618  	\\ \hline  
		\end{tabular}
\caption {Same as Table VI but for showers initiated by S-II nuclei (QGSJet-II-04 model).}
	\end{center}
\end{table*}

 \begin{figure}[!htbp]
	\centering
	\includegraphics[trim=0.6cm 0.6cm 0.6cm 0.6cm, scale=0.8]{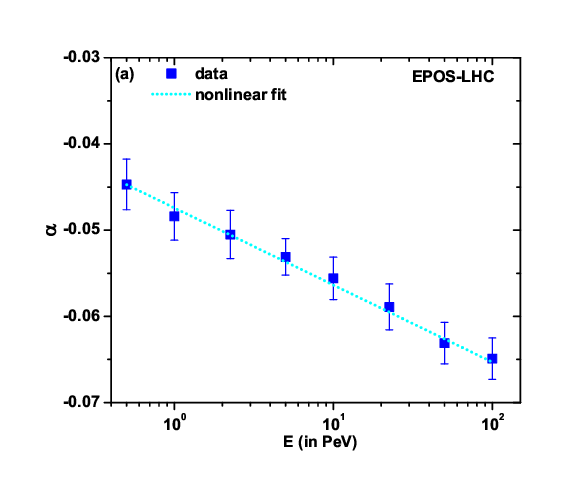}
	\includegraphics[trim=0.6cm 0.6cm 0.6cm 0.6cm, scale=0.8]{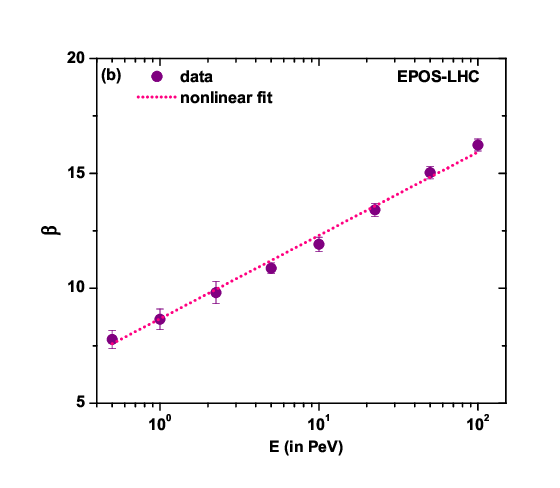}
	\caption{Fig.~a: $\alpha$ parameters estimated at eight different CR energies $E$ for showers initiated by S-I nuclei. Fig.~b: $\beta$ parameters estimated at eight different CR energies $E$ for showers initiated by S-I nuclei.}
\end{figure}

\begin{table*}[!htbp]
	\begin{center}
	\begin{tabular}{|p{0.16\textwidth}| p{0.16\textwidth}|  p{0.16\textwidth} | }
			\cline{1-3}
			$~~~~~~~~~a_1$& {\bf~~~~~~~~~$b_1$} & {\bf~~~~~~~~~$c_1$} \\ \hline
			0.1991 $\pm$ 0.0002 & -0.4020 $\pm$ -0.0002& -0.0475 $\pm$ 0.0004	\\ \hline\hline
			$~~~~~~~~~a_2$& {\bf~~~~~~~~~$b_2$} & {\bf~~~~~~~~~$c_2$} \\ \hline
			0.5841 $\pm$ 0.0373 & 0.4065 $\pm$ 0.0376 & 8.6743 $\pm$ 0.0867	\\ \hline 
		\end{tabular}
\caption{The fitted parameters in Eq.~(9) and (10) for showers initiated by S-I (EPOS-LHC) nuclei.}
	\end{center}
\end{table*}

\begin{figure}[!htbp]	
	\centering
	\includegraphics[trim=0.6cm 0.6cm 0.6cm 0.6cm, scale=0.8]{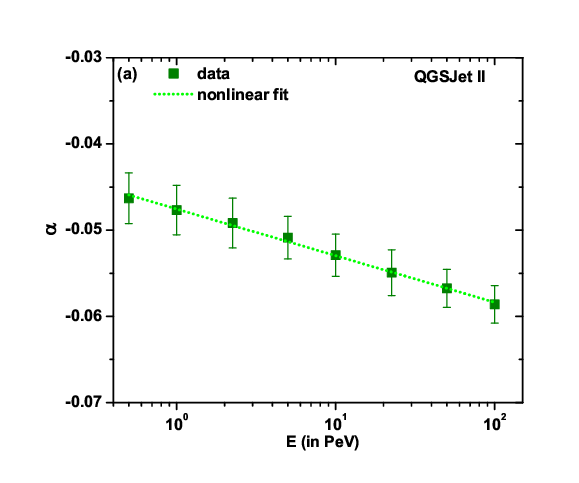}
	\includegraphics[trim=0.6cm 0.6cm 0.6cm 0.6cm, scale=0.8]{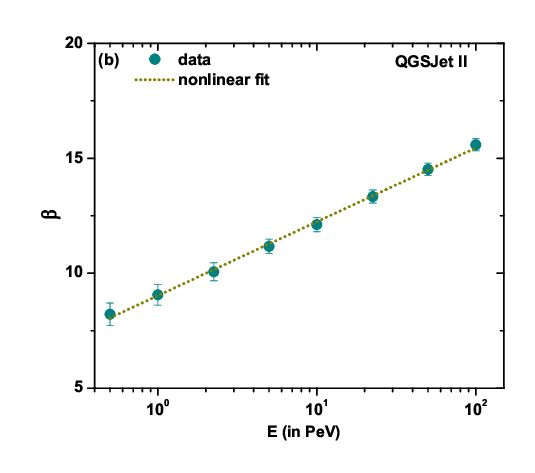}
	\caption{Fig.~a: $\alpha$ parameters estimated at eight different CR energies $E$ for showers initiated by S-II nuclei. Fig.~b: $\beta$ parameters estimated at eight different CR energies $E$ for showers initiated by S-II nuclei.}
\end{figure}	

\begin{table*}[!htbp]
	\begin{center}
		\begin{tabular}{|p{0.16\textwidth}| p{0.16\textwidth}|  p{0.16\textwidth} | }
			\cline{1-3}
			$~~~~~~~~~a_1$& {\bf~~~~~~~~~$b_1$} & {\bf~~~~~~~~~$c_1$} \\ \hline
			0.2244 $\pm$ 0.0001 & -0.4511 $\pm$ -0.0001 & -0.0476 $\pm$ -0.0002	\\ \hline\hline
			$~~~~~~~~~a_2$& {\bf~~~~~~~~~$b_2$} & {\bf~~~~~~~~~$c_2$} \\ \hline
			0.5738 $\pm$ 0.0354 & 0.2476 $\pm$ 0.0248   & 9.0276 $\pm$ 0.0559	\\ \hline 
		\end{tabular}
\caption{Same as Table VIII but for showers initiated by S-II (QGSJet-II-04) nuclei.}
	\end{center}
\end{table*}
  
\begin{figure}[!htbp]
	\centering
	\includegraphics[trim=0.6cm 0.6cm 0.6cm 0.6cm, scale=0.8]{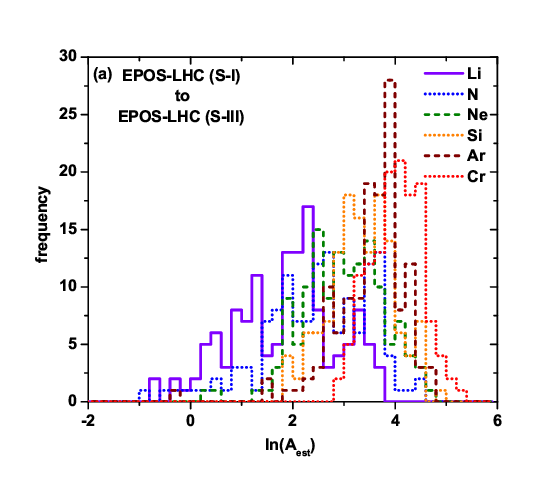}
	\includegraphics[trim=0.6cm 0.6cm 0.6cm 0.6cm, scale=0.8]{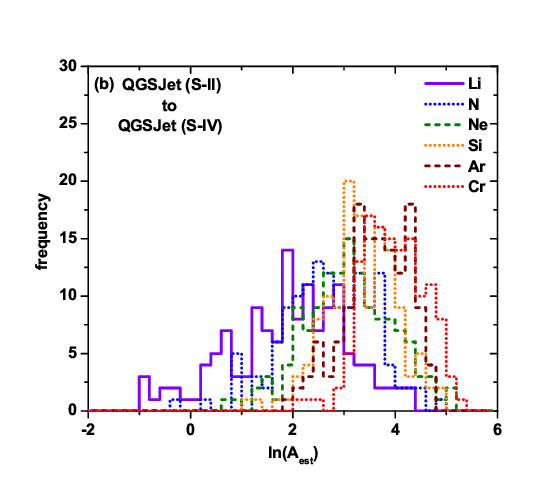}
	\caption{Fig.~a: distributions of estimated values of $\ln_{A_{\text{est}}}$ for S-III (EPOS-LHC) showers with parameters determined by the S-I (EPOS-LHC) showers. Fig.~b: same as Fig.~a but for showers initiated by S-IV (QGSJet-II-04) nuclei with parameters determined by the S-II (QGSJet-II-04) showers.}
\end{figure}
 
\section{Results and discussions}             
\subsection{Correlation of $\eta_{\rho}$ with the lateral shower age $s_{\text{av}}$}
The generated showers in samples S-I and S-II manifest a condition for both the EPOS-LHC and QGSJet-II-04 models, which suggests a correlation between $s_{\text{av}}$ and $\eta_{\rho}$. Such a correlation between two observables can be parametrized in a straightforward exponential form irrespective of the high-energy hadronic interaction models as  

 \begin{equation}
s_{\text{av}}(\eta_{\rho}) = a+b \cdot exp(-\eta_{\rho}/c)
\end{equation}        

For each simulated p-initiated shower, say, in sample S-I (EPOS-LHC), a pair of parameters ($s_{\text{av}},\eta_{\rho}$) was obtained. We have then combined all the pairs of such parameters obtained from the analysis of a total of 340 p-initiated simulated showers of the sample S-I (EPOS-LHC) for eight different primary energies in a figure (Fig.~3a), which is displaying a variation of $s_{\text{av}}$ with $\eta_{\rho}$. For the remaining nuclei, He, C, O, Mg, S, Ca and Fe in S-I, the dependence of $s_{\text{av}}$ on $\eta_{\rho}$ are shown in Fig.~3b to Fig.~3h. All these variations of $s_{\text{av}}$ with $\eta_{\rho}$ in Fig.~3 can be described by the parametrization in Eq.~(7). For each type of primary shower in S-I (EPOS-LHC), the parameters $a$, $b$ and $c$ are found by making an exponential fit according to Eq.~(7) of $s_{\text{av}}$ versus $\eta_{\rho}$ data. As a result, eight sets of fit parameters ($a$, $b$ and $c$) can be obtained for eight primary particle types. Finally, the correlation between $s_{\text{av}}$ and $\eta_{\rho}$ in Eq.~(7) is also used to fit all the 2720 number of $s_{\text{av}}$ versus $\eta_{\rho}$ data in S-I (EPOS-LHC) together. The combined scattered plot of $s_{\text{av}}$ versus $\eta_{\rho}$ data, as well as the fitted curve, is shown in Fig.~4. The parameters, $a$, $b$ and $c$ obtained from the fits in Fig.~3 and Fig.~4 are shown in Table III. On the other hand, Fig.~5, Fig.~6 and Table IV together obtained from the analysis of simulated showers in sample S-II (QGSJet-II-04) display the analogous results of Fig.~3, Fig.~4 and Table III. It is noteworthy to mention that the nature of variation of $s_{\text{av}}$ with the variation of $\eta_{\rho}$ is effectively independent of the hadronic interaction models that were used in the present simulation.\\ 

The $\eta_{\rho}$ values experience statistical fluctuations with the type of primary species and CR energy. This may cause a dispersion in the correlation of $s_{\text{av}}$ with $\eta_{\rho}$ using simulations. For a specific case with $E=50$~PeV, our analysis has found a larger error in $\eta_{\rho}$ with a dispersion, $\frac{\Delta{\eta_{\rho}}}{\eta_{\rho}}\times 100\%\approx \pm 12.3\%$ for the primary p. Such a dispersion has declined to $\pm 6.2\%$ if it corresponded to Fe showers. Our simulation gives a dispersion range, $\approx \pm{(4 - 11)}\%$ in $\eta_{\rho}$ for Fe showers, if the simulation corresponded to a primary energy range, $0.5 - 100$~PeV. The main reason for higher errors of $\eta_{\rho}$ at lower primary energies is the large shower-to-shower fluctuations in electron densities, particularly around $310$~m. Radial distance beyond $300$~m, a considerable fall in electron density may contribute a significant error to $\eta_{\rho}$. As the lateral spread of p or lighter nuclei-initiated showers is smaller than that of the Fe-initiated showers, the error of $\eta_{\rho}$ is expected to be higher for showers initiated by p or lighter nuclei.\\

The dispersion found in $\eta_{\rho}$ should affect the value of $s_{\text{av}}$ via their correlation in Eq.~(7). We have seen that for $50$~PeV proton-initiated showers the average values of $\eta_{\rho}$ and $\Delta{\eta_{\rho}}$ were $237.42$ and $\pm{29.3}$. Using these values, together with $a,b,c$ from Table III for proton primary in Eq.~(7), the average values of $s_{\text{av}}$ and $\Delta{s_{\text{av}}}$ are $\approx{1.13}$ and $\pm{0.04}$. We notice that these results on $s_{\text{av}}$ and $\Delta{s_{\text{av}}}$ agree very well with those obtained from LAP, as mentioned in Sect. III. Hence, using the new observable, $\eta_{\rho}$ in predicting the shower age parameter looks practicable.\\

\subsection{Applicability of Eq.~(7) to Group II showers}
To demonstrate the applicability of the present method via Eq.~(7) to Group II showers, we would exploit explicitly the combined fit parameters obtained from the analysis of Group I showers in Group II. For S-III (EPOS-LHC) and S-IV (QGSJet-II-04) samples of showers in Group II, the parameter $s_{\text{av}}$ is determined from the average of several LAPs in the radial distance range, $45\textendash{310}$~m, as done before. Further, the parameter $\eta_{\rho}$ is also determined for them. In order to test the validity of our analysis algorithm, we analyzed Group II:EPOS-LHC(QGSJet-II-04) showers by exploiting the parameters determined by Group I:EPOS-LHC(QGSJet-II-04) showers. To test the high-energy hadronic interaction model dependency, we have, however, analyzed the Group II:QGSJet-II-04(EPOS-LHC) showers with the parameters determined by the Group I:EPOS-LHC(QGSJet-II-04) showers.\\

Thus, first using the observable $\eta_{\rho}$ of each shower in the S-III (EPOS-LHC) sample, we would estimate their corresponding $s_{\text{est}}$ values with the help of Eq.~(7) together with the combined fit parameters $a,b,c$ borrowed from Table III (highlighted $a,b,c$ only) for the S-I (EPOS-LHC) sample. We compared the shower age, $s_{\text{est}}$ predicted by the relation (7), with $s_{\text{av}}$ of showers generated by nuclei of the S-III (EPOS-LHC) sample. The results are displayed in Fig.~7a, which reveal that the $s_{\text{av}}$ parameter can be expressed as a function of $s_{\text{est}}$ by a linear equation, $s_{\text{av}}=a_{1}~{s_{\text{est}}}+s_{0}$ with fitted parameters ($a_{1}=0.97\pm{0.004},s_{0}=0.032\pm{0.005}$) for the Group II nuclei with the S-III (EPOS-LHC) sample, respectively. We repeat it by analyzing S-IV (QGSJet-II-04) showers with the parameters determined by S-II (QGSJet-II-04) showers. Likewise, we depict the obtained results in Fig.~7b, and the corresponding fitted parameters take the values as ($a_{1}=0.98\pm{0.005},s_{0}=0.021\pm{0.006}$).

Secondly, we have analyzed the Group II:QGSJet-II-04(EPOS-LHC) showers with the parameters determined by the Group I:EPOS-LHC(QGSJet-II-04) showers. These results are displayed in Fig.~8a and Fig.~8b, and the corresponding fitted parameters are ($a_1=0.98\pm{0.005},s_{0}=0.022\pm{0.006}$) and ($a_1=0.98\pm{0.004},s_{0}=0.027\pm{0.005}$). Moreover, we further computed the difference between the two age parameters, $s_{\text{av}}$ and $s_{\text{est}}$. Finally, in Fig.~7c and Fig.~7d, we have presented a pair of histograms of these differences, $\delta{s}=s_{\text{av}}-s_{\text{est}}$ for showers initiated by Group II nuclei with S-III(EPOS-LHC) and S-IV(QGSJet-II-04) samples, respectively. The Gaussian distribution function fits the histogram (Fig.~7c) with a parameter set, $\mu = 0.0$ and $\sigma = 0.0056$, and the histogram (Fig.~7d) with $\mu = -0.0017$ and $\sigma = 0.0054$. On the other hand, the histogram fits of Fig.~8c and d picked the values as $\mu = -0.0017,\sigma = 0.0054$ and $\mu = 0.001,\sigma = 0.005$, respectively. It has been thus noticed that the relation of $s_{\text{av}}$ or $s_{\text{est}}$ with $\eta_{\rho}$ obtained from the present analysis clearly shows insignificant model dependence.\\

In the above, we have parametrized $s_{\text{est}}$ as a function of $\eta_{\rho}$ of Group II showers using a set of combined fit parameters $a$, $b$ and $c$, derived from Group I showers for both EPOS-LHC and QGSJet-II-04 hadronic interaction models. It should be, however, mentioned how the uncertainties in combined fit parameters $a$, $b$, and $c$ given in Table III (EPOS-LHC) would lead to cast uncertainties in $s_{\text{est}}$ in Table V. The error propagation procedure puts an estimate as,\\

$\Delta{s_{\text{est}}}=\Big[\Delta a^2 + \frac{exp({-{2}\eta_{\rho}/{c})\cdot \{b^2 (c^2 \Delta \eta_{\rho}^2+  \Delta c^2 \eta_{\rho}^2)+ \Delta b^2 c^2\} }}{c^4}\Big]^{1/2}$;\\

with $\eta_{\rho}$ as well as $\Delta{\eta_{\rho}}$ come from Group II showers. In Table V, we have shown $s_{\text{est}}\pm{\Delta{s_{\text{est}}}}$, corresponding to $\eta_{\rho}\pm{\Delta{\eta_{\rho}}}$ of Group II showers generated by EPOS-LHC model only. Though not shown here, our analysis found no noticeable difference in these results using the QGSJet-II-04 model.   

\subsection{An estimate of cosmic ray mass $A$ extracted from $\eta_{\rho}$}

We know that $s$ or $s_{\text{av}}$ takes the highest value for the CR nuclei $A=56$ and the lowest value for $A=1$ at a particular $E$ or a particular pair of EAS observables ($N_{e}, N_{\mu}$) in simulations [17,29]. Commonly, the comparison of experimentally measured variations of $s$ or $s_{\text{av}}$ versus $E$ or versus $N_{e}$ or versus the combined observables $N_{e}, N_{\mu}$ with simulations has been exploited to study the change of CR mass composition across the knee or ankle [17,29-30]. Here, we observe, though in simulated data, that $\eta_{\rho}$ as a new EAS observable gets the higher value for the CR species $A=1$ and the lowest value for $A=56$ at a particular $E$ (see Fig.~9a or Table V with EPOS-LHC model, and Fig.~9b only with QGSJet-II-04 model). A direct measure of $A$ could be achieved if a correlation between $\eta_{\rho}$ and $A$ is established. Based on a set of simulated showers, a unique correlation is obtained to estimate the CR mass $A$ by exploiting the EAS observable $\eta_{\rho}$ alone. At a specific CR energy, the observable $\eta_{\rho}$ accounts the variations in $A$ among different CR species. The energy dependence of $A$ is directly regulated by the two parameters $\alpha$ and $\beta$. The correlation is then given by, 

\begin{equation}
	A = exp{(\alpha\cdot\eta_{\rho} + \beta)}
	\end{equation}		
where, $A$ is expressed in atomic mass unit (amu). 

To amend the parameters $\alpha$ and $\beta$, we use the following analytical models (see Eq.~(9) and Eq.~(10)) that have been derived by fitting the Eq.~(8) to the data of $A$ versus $\eta_{\rho}$ obtained from the analysis of showers in samples S-I (EPOS-LHC) and S-II (QGSJet-II-04). The parametrized $\alpha$ and $\beta$ take the following dependencies on the CR energy. 
	\begin{equation}
	\alpha{(E)} = a_{1}\cdot ln(E^2) + b_{1}\cdot ln(E)+c_{1}  
	\end{equation}
	\begin{equation}
	\beta{(E)}  = a_{2}\cdot ln(E^2) + b_{2}\cdot ln(E)+c_{2}
	\end{equation}

The parameters $\alpha$ and $\beta$ are so obtained from the fit procedure using Eq.~(8) to the variations of $A$ with $\eta_{\rho}$ for showers in S-I (EPOS-LHC) and S-II (QGSJet-II-04) samples, are listed in Table VI and Table VII. These variations between $A$ and $\eta_{\rho}$ are already shown in Fig.~9a and b. The corresponding fitted values of $\alpha$ and $\beta$ for different primary energies are shown in Table VI (S-I) and Table VII (S-II). The variations of fitted $\alpha$ and $\beta$ with the CR energy are parametrized nicely by Eq.~(9) and Eq.~(10) and are shown in Fig.~10a and b for the S-I (EPOS-LHC) sample and in Fig.~11a and b for the S-II (QGSJet-II-04) sample. The six parameters ($a_1$, $b_1$, $c_1$) and ($a_2$, $b_2$, $c_2$) are included in Table VIII and Table IX.

In order to validate the analytical modeling for the CR mass $A$ via Eq.~(8) to Group-II showers in S-III (EPOS-LHC) sample, we would borrow the parameters ($a_1$, $b_1$, $c_1$) and ($a_2$, $b_2$, $c_2$) obtained from the analysis of the S-I sample in Table VIII, and use them in S-III (EPOS-LHC) showers to obtain $\alpha$ and $\beta$ corresponding to their CR energies. For the S-III (EPOS-LHC) sample of simulated showers, the parameter $\eta_{\rho}$ is already known from Table V, corresponding to the EPOS-LHC model. Now, using the parameter $\eta_{\rho}$ of each shower in the sample S-III (EPOS-LHC) and the combined parameters $\alpha$ and $\beta$, the estimated masses $A_{\text{est}}$ for the species in S-III (EPOS-LHC) can be predicted with the help of Eq.~(8). A reasonably more precise estimate of the primary CR mass $A_{\text{est}}$ and a better separation of the distributions from individual types of primary species can be achieved if the resolution in $A_{\text{est}}$ determination is expressed in $\ln{A_{\text{est}}}$. The distributions of $\ln{A_{\text{est}}}$ analyzed from the S-III (EPOS-LHC) showers with the parameters determined by the S-I (EPOS-LHC) showers are depicted in Fig.~12a. Six different types of species in the S-III sample are included. Similarly in Fig.~12b, the distributions of $\ln{A_{\text{est}}}$ analyzed from the S-IV (QGSJet-II-04) showers with the parameters determined by the S-II (QGSJet-II-04) showers are also shown. The mean ($\mu_{(\ln{A_{\text{est}}})}$) and standard deviation ($\sigma_{(\ln{A_{\text{est}}})}$) of the distributions for S-III (EPOS-LHC) and S-IV (QGSJet-II-04) showers are given in Table X and Table XI, respectively. To see a certain level of high-energy hadronic interaction model dependency of $\ln{A_{\text{est}}}$ distributions through the characteristic parameters $\mu_{(\ln{A_{\text{est}}})}$ and $\sigma_{(\ln{A_{\text{est}}})}$, we have, however, analyzed the Group II:QGSJet-II-04(EPOS-LHC) showers with the parameters determined by the Group I:EPOS-LHC(QGSJet-II-04) showers. In the later cases, we have just shown these final characteristic parameters via Table XII and Table XIII from the analysis. Thus, the resulting estimates for $\ln{A_{\text{est}}}$, shown in Fig.~12 or Table X to Table XIII, reveal the efficiency of the present method to identify heavy nuclei (e.g., Cr) from light CR primaries (e.g., Li). Alternatively, in $\ln{A_{\text{est}}}$, the resolution is clearly comparable for heavy and light primaries.

\begin{table}[!htbp]
	\begin{tabular}{|l|l|l|l|}
		\hline
		PCR             & $\ln{(A_{actual})}$  & $\mu_{(\ln{A_{est}})}$      & $\sigma_{(\ln{A_{est}})}$     \\ \hline
		$Li_{3}^{7}$    & 1.946  &1.939 $\pm$0.143 & 1.069 $\pm$0.147 \\ \hline
		$N_{7}^{14}$    & 2.639  &2.676 $\pm$0.115 & 1.031 $\pm$0.119 \\ \hline
		$Ne_{10}^{20}$  & 2.996  &3.006 $\pm$0.067 & 0.834 $\pm$0.068 \\ \hline
		$Si_{14}^{28}$  & 3.332  &3.372 $\pm$0.054 & 0.639 $\pm$0.055 \\ \hline
		$Ar_{18}^{40}$  & 3.689  &3.708 $\pm$0.066 & 0.501 $\pm$0.066 \\ \hline
		$Cr_{24}^{52}$  & 3.952  &4.048 $\pm$0.036 & 0.530 $\pm$0.035 \\ \hline
	\end{tabular}
	\caption{The resulting values of the mean $\mu_{(\ln{A_{\text{est}}})}$ and standard deviation $\sigma_{(\ln{A_{est}})}$ from the Gaussian fit of the distributions of estimated $\ln{A_{\text{est}}}$) of S-III (EPOS-LHC) showers with parameters determined by S-I (EPOS-LHC) showers.}
\end{table}

\begin{table}[!htbp]
	\begin{tabular}{|l|l|l|l|}
		\hline
		PCR   & $ln(A_{actual})$   & $\mu_{(lnA_{est})}$ & $\sigma_{(lnA_{est})}$     \\ \hline
		$Li_{3}^{7}$    & 1.946  &2.082 $\pm$0.106 & 1.116 $\pm$0.108 \\ \hline
		$N_{7}^{14}$    & 2.639  &2.766 $\pm$0.063 & 0.888 $\pm$0.063 \\ \hline
		$Ne_{10}^{20}$  & 2.996  &3.091 $\pm$0.042 & 0.865 $\pm$0.043 \\ \hline
		$Si_{14}^{28}$  & 3.332  &3.356 $\pm$0.064 & 0.672 $\pm$0.063 \\ \hline
		$Ar_{18}^{40}$  & 3.689  &3.717 $\pm$0.033 & 0.656 $\pm$0.033 \\ \hline
		$Cr_{24}^{52}$  & 3.952  &3.948 $\pm$0.049 & 0.643 $\pm$0.030 \\ \hline
	\end{tabular}
	\caption{The resulting values of the mean $\mu_{(\ln{A_{\text{est}}})}$ and standard deviation $\sigma_{(\ln{A_{est}})}$ from the Gaussian fit of the distributions of estimated $\ln{A_{\text{est}}}$) of S-IV (QGSJet-II-04) showers with parameters determined by S-II (QGSJet-II-04) showers.}
\end{table}

\newpage

\begin{table}[!htbp]
	\begin{tabular}{|l|l|l|l|}
		\hline
		PCR             & $ln(A_{actual})$  & $\mu_{(lnA_{est})}$      & $\sigma_{(lnA_{est})}$     \\ \hline
		$Li_{3}^{7}$    & 1.946  &2.567 $\pm$0.097 & 1.095 $\pm$0.099 \\ \hline
		$N_{7}^{14}$    & 2.639  &3.319 $\pm$0.059 & 0.794 $\pm$0.081 \\ \hline
		$Ne_{10}^{20}$  & 2.996  &3.512 $\pm$0.078 & 0.772 $\pm$0.071 \\ \hline
		$Si_{14}^{28}$  & 3.332  &3.907 $\pm$0.060 & 0.573 $\pm$0.060 \\ \hline
		$Ar_{18}^{40}$  & 3.689  &4.301 $\pm$0.032 & 0.568 $\pm$0.061 \\ \hline
		$Cr_{24}^{52}$  & 3.952  &4.531 $\pm$0.069 & 0.451 $\pm$0.042 \\ \hline
	\end{tabular}
	\caption{The resulting values of the mean $\mu_{(\ln{A_{\text{est}}})}$ and standard deviation $\sigma_{(\ln{A_{est}})}$ from the Gaussian fit of the distributions of estimated $\ln{A_{\text{est}}}$) of S-III (EPOS-LHC) showers with parameters determined by S-II (QGSJet-II-04) showers.}
\end{table}

\begin{table}[!htbp]
	\begin{tabular}{|l|l|l|l|}
		\hline
		PCR             & $ln(A_{actual})$  & $\mu_{(lnA_{est})}$      & $\sigma_{(lnA_{est})}$     \\ \hline
		$Li_{3}^{7}$    & 1.946  &1.739 $\pm$0.101 & 1.093 $\pm$0.102 \\ \hline
		$N_{7}^{14}$    & 2.639  &2.280 $\pm$0.113 & 0.975 $\pm$0.118 \\ \hline
		$Ne_{10}^{20}$  & 2.996  &2.595 $\pm$0.119 & 0.955 $\pm$0.125 \\ \hline
		$Si_{14}^{28}$  & 3.332  &2.885 $\pm$0.054 & 0.629 $\pm$0.054 \\ \hline
		$Ar_{18}^{40}$  & 3.689  &3.313 $\pm$0.089 & 0.751 $\pm$0.091 \\ \hline
		$Cr_{24}^{52}$  & 3.952  &3.509 $\pm$0.077 & 0.751 $\pm$0.079 \\ \hline
	\end{tabular}
	\caption{The resulting values of the mean $\mu_{(\ln{A_{\text{est}}})}$ and standard deviation $\sigma_{(\ln{A_{est}})}$ from the Gaussian fit of the distributions of estimated $\ln{A_{\text{est}}}$) of S-IV (QGSJet-II-04) showers with parameters determined by S-I (EPOS-LHC) showers.}
\end{table}

\begin{figure}[!htbp]
	\centering
	\includegraphics[trim=0.6cm 0.6cm 0.6cm 0.6cm, scale=0.8]{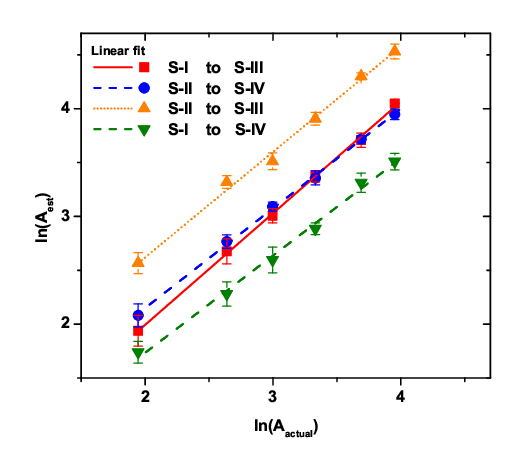}
	\caption{Graphical representation of the numerals in Table X to XIII, showing the estimated ($\ln{A_{\text{est}}}$) values corresponding to their actual values ($\ln{A_{\text{actual}}}$).}
\end{figure}

The mean values of $\ln{(A_{est})}$ i.e., $\mu_{(\ln{A_{est}})}$ are plotted against $\ln{(A_{actual})}$, obtained from Table X to Table XIII in Fig.~13, and then fitted with the straight line $\ln{(A_{est})} = m \cdot \ln{(A)} + c $. An insignificant difference between $\ln{(A_{est})}$ i.e. $\mu_{(\ln{A_{est}})}$ and $\ln{(A_{actual})}$ for the cases EPOS-LHC (S-I) to EPOS-LHC (S-III) (solid-red line in Fig.~13) and QGSJet-II-04 (S-II) to QGSJet-II-04 (S-IV) (dash-blue line) indicates the validity of our analysis algorithm utilized in the work. Contrary to that, the two peripheral straight lines (green and amber) in Fig.~13 corresponding to EPOS-LHC (S-I) to QGSJet-II-04 (S-IV), and QGSJet-II-04 (S-II) to EPOS-LHC (S-III) cases indicate a certain level of high energy interaction model dependency of our CR mass determination method.

\begin{figure}[!htbp]
	\centering
	\includegraphics[trim=0.6cm 0.6cm 0.6cm 0.6cm, scale=0.8]{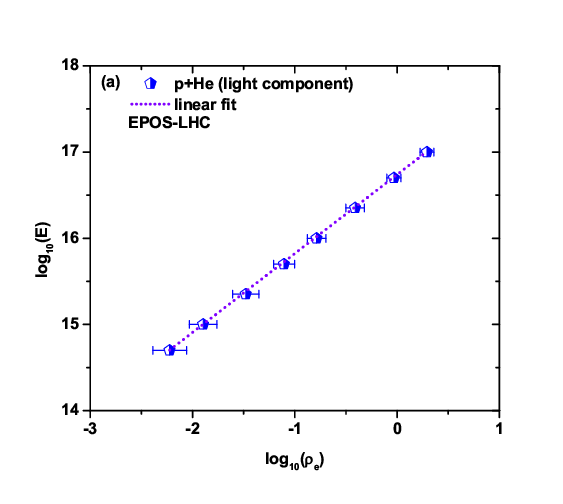}
        \includegraphics[trim=0.6cm 0.6cm 0.6cm 0.6cm, scale=0.8]{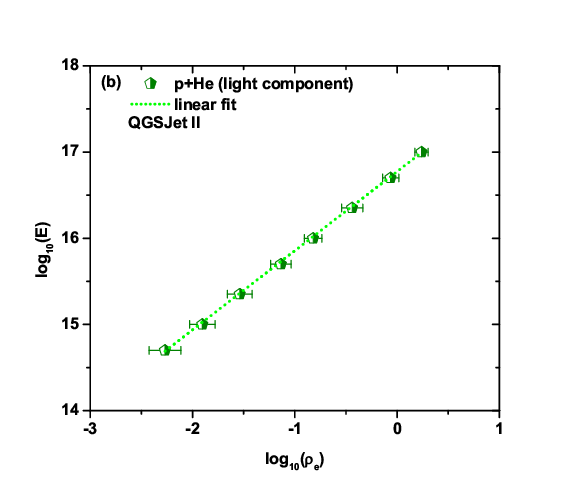}   
	\caption{Correlation of the CR energy of S-I (EPOS-LHC) showers and average density of EAS charged particles ($e+\mu$) at a core distance of $310$~m.}
\end{figure}
\begin{figure}[!htbp]
	\centering
	\includegraphics[trim=0.6cm 0.6cm 0.6cm 0.6cm, scale=0.8]{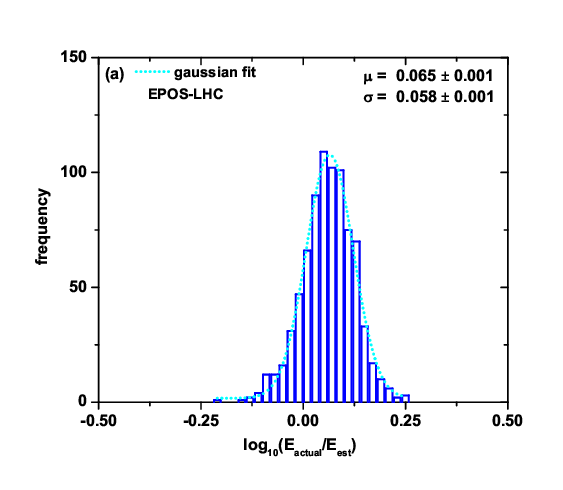}
        \includegraphics[trim=0.6cm 0.6cm 0.6cm 0.6cm, scale=0.8]{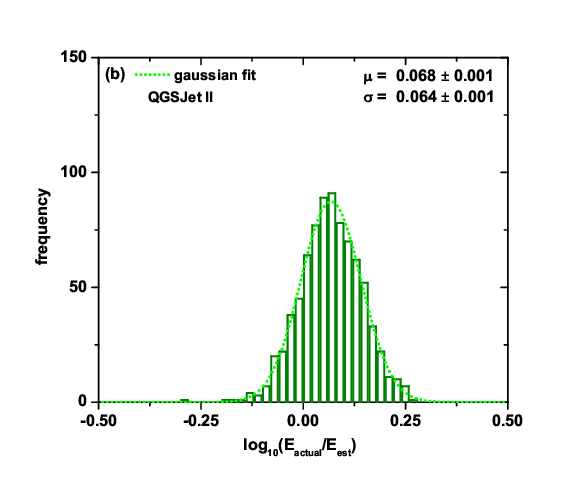}
	\caption{Distributions of $log_{10}{(E_{\text{actual}}/E_{\text{est}})}$ for showers initiated by S-III (EPOS-LHC) nuclei. The values of the mean and standard deviation of the distribution are shown.}
\end{figure}

The correlation in Eq.~(8) is applied to estimate the average CR mass,  requiring a measure of the CR energy $E$. Some earlier simulations determined $E$ in the EeV range by using the estimators as either the flux density of EAS charged particles ($\rho_{e+\mu}$) or the flux density of only muons ($\rho_{\mu}$) both measured at a core distance of $600$~m [48]. Our simulation study, however, put some effort just to examine whether the selected fixed CR energies in the work could be estimated using the correlation between $E$ and the local density of electrons ($\rho_{e}$) around $310$~m radial distance and at least for the light component (proton and helium) air showers by the following,
\begin{equation}
log_{10}{(E)} = A+B~log_{10}{(\rho_{e}(r=310~{\text{m}}))}
\end{equation}
Analyses of simulated data find $A=16.74\pm{0.01}$ and $B=0.92\pm{0.01}$ from the EPOS-LHC model whereas using the QGSJet-II-04 these values are $A=16.77\pm{0.02}$ and $B=0.92\pm{0.01}$ respectively. The relationship (Eq.~(11)) for the light component (proton and helium) obtained in this study can be used for energy estimations of S-III and S-IV nuclei-initiated showers. The average differences between the actual and estimated CR energies are $log_{10}{(E_{\text{actual}}/E_{\text{est}})}\simeq 0.065$ and $\simeq 0.068$ respectively from the EPOS-LHC and QGSJet-II models. In Fig.~14 and 15, our analyses are presented, considering both the interaction models.\\

Finally, the CR mass sensitivity power of the new observable $\eta_{\rho}$ is tested using a well-known statistical parameter {\lq the figure of merit (FOM)\rq}. The FOM could determine the quality of the CR mass estimator $\eta_{\rho}$. The FOM is a very reliable statistical indicator for Gaussian distributions. We estimate the FOM for the first two classes (e.g. $c_{1}=\text{p}$ and $c_{2}=\text{Fe}$ classes) of Group-I nuclei and then for the second two classes (e.g. $c_{1}=\text{Li}$ and $c_{2}=\text{Cr}$ classes) of Group-II. The standard form of the FOM is written by
 
\begin{equation}
\text{FOM} =  \frac{ \left|\mu[\eta_{\rho_{\text{c}_1}}] - \mu[\eta_{\rho_{\text{c}_2}}]\right| }{\sqrt{\sigma^2[\eta_{\rho_{\text{c}_1}}]+\sigma^2[\eta_{\rho_{\text{c}_2}}]}}
\end{equation}

, where $\mu[\eta_{\rho_{\text{c}_1}}]$ and $\sigma^2[\eta_{\rho_{\text{c}_1}}]$ are the mean value and the variance of the distribution of the observable $\eta_{\rho}$ for $\text{c}_1$ class showers while $\mu[\eta_{\rho_{\text{c}_2}}]$ and $\sigma^2[\eta_{\rho_{\text{c}_2}}]$ for $\text{c}_2$ class showers respectively. The FOMs for the observable $\eta_{\rho}$ take $1.98$ and $2.36$, respectively, corresponding to the EPOS-LHC and QGSJet-II models from the first two classes of simulated showers at a fixed CR energy $50$~PeV. The FOMs are $1.77$ and $1.60$ from the second two classes of simulated showers generated at the energy $32$~PeV, corresponding to the EPOS-LHC and QGSJet-II models. 

\section{Conclusions}
In the present analysis using a new method, we mainly focused on the observable $\eta_{\rho}$ for extracting the CR mass information and have drawn the following conclusions.

It was demonstrated that the new observable $\eta_{\rho}$ has maintained a good correlation with the lateral shower age $s_{\text{av}}$. The relation between $s_{\text{av}}$ and $\eta_{\rho}$ for Group I showers is found very accurate and reliable in predicting lateral shower age $s_{\text{est}}$ of EASs initiated by a new CR series of nuclei in Group II. The predicted lateral shower age $s_{\text{est}}$ simply using the observable $\eta_{\rho}$ agrees nicely with that obtained from the existing method of averaging several LAPs of an EAS in the radial distance range $45-310$~m (shown in Table V and also in Fig.~7, and Fig.~8). As $\eta_{\rho}$ is found a complementary one to the lateral shower age $s_{\text{av}}$ or $s_{\text{est}}$ of an EAS, it seems evident that the observable be effective to characterize EASs. Our analyses agree with some well-established facts that lateral shower age rises with the CR mass $A$ at a fixed primary energy and decreases with the CR energy $E$ for a particular CR species (Table V and elsewhere). The observable $\eta_{\rho}$ also retains such dependencies on $A$ and $E$ but in a converse manner.

Some of the features of the lateral shower age known from previous studies revealed that the parameter and its fluctuation showed sensitivity to the nature of the shower-initiating primary CR particle. The lateral shower age $s_{\text{est}}$ estimated by exploiting the new observable $\eta_{\rho}$ may presumably improve the accuracy of the measurement of the CR mass composition by air shower experiments.

The work mainly aimed to exploit the observable $\eta_{\rho}$ as an effective estimator of the primary CR mass. It has been demonstrated that the method can estimate the average CR masses from simulated EAS data, if $\eta_{\rho}$ and $E$ are measured with average accuracy (Fig.~12 and Table X to Table XIII).

The present analyses rely on air shower simulations, and therefore, validating the results obtained with two hadronic interaction models, EPOS-LHC and QGSJet-II-04, is very appropriate to gain a greater insight into differences, if any. It has been thus noticed that relation of $s_{\text{av}}$ or $s_{\text{est}}$ with $\eta_{\rho}$, obtained from the present analysis clearly showing a little model dependence. However, a certain level of high energy interaction model dependency of the CR mass determination from the present method has been noticed. A clear indication of that feature could be realized by the interpretation of the departure of the two peripheral straight lines (green and amber) in Fig.~13 from the central ones.

In this paper, we have proposed a new observable, studied its characteristics, and discussed its application for simulated air shower data. We noticed that $\eta_{\rho}$ values experienced an adequate dispersion range from their averages owing to significant shower-to-shower statistical uncertainties in $\rho_{\text{e}}$, particularly at large radial distances. A ground array equipped with densely packed scintillation detectors only covering up to $\simeq{350}$~m is suitable to determine $\eta_{\rho}$, and of course, the core locations of EASs (would provide very accurate radial distance measurement) more accurately.

It is worth mentioning that the present analysis has been restricted by some limiting factors in connection with the present computational capabilities  for simulations. Accordingly, we planned to go along with a modest total of $2\times{3530}$ vertical showers generated at different fixed primary energies in the range of $0.5-128$~PeV at the default KASCADE location in CORSIKA. It would be crucial to formulate the main attribute of the work to inclined simulated EASs with arbitrary energy in the range $0.5-130$~PeV, and also to examine a specific surface detector array response in future. Applying the basic aspect and the scheme of the method to a CR composition analysis using observed electron density data from a closely packed array of scintillation detectors [49,50] would be an exciting project.

\section*{Acknowledgment}
Authors acknowledge the financial support from the SERB, Department of Science and Technology (Govt. of India) under Grant No. EMR/2015/001390.

\end{document}